\def\final{1}
\def\lncs{0}
\def\lncsshort{0}

\ifnum\lncs=1
	\documentclass[envcountsame]{llncs}
\else
	\documentclass[11pt]{article}
	
	\usepackage{amsthm}
	\usepackage{enumerate}
	\usepackage{enumitem}
	\usepackage[margin=1.1in]{geometry}
	\usepackage{microtype}
	\usepackage{kpfonts}
		\DeclareMathAlphabet{\mathsf}{OT1}{cmss}{m}{n}
		\SetMathAlphabet{\mathsf}{bold}{OT1}{cmss}{bx}{n}

	\usepackage{multirow, tabularx}
	\usepackage{color}
	\definecolor{DarkGreen}{rgb}{0.15,0.5,0.15}
	\definecolor{DarkRed}{rgb}{0.6,0.2,0.2}
	\definecolor{DarkBlue}{rgb}{0.15,0.15,0.55}
	\definecolor{DarkPurple}{rgb}{0.4,0.2,0.4}

	\usepackage[pdftex]{hyperref}
	\hypersetup{
		   linktocpage=true,
 		   colorlinks=true,				
 		   linkcolor=DarkBlue,		
 		   citecolor=DarkBlue,	
 		   urlcolor=DarkBlue,		
	}
	
	\newcolumntype{Y}{>{\centering\arraybackslash}X}
	
\fi

\usepackage{algorithm, algorithmic}
\usepackage{amsmath, amssymb}
\usepackage{framed}
\usepackage{verbatim}
\usepackage{color}
	\definecolor{DarkGreen}{rgb}{0.15,0.5,0.15}
	\definecolor{DarkRed}{rgb}{0.6,0.2,0.2}
	\definecolor{DarkBlue}{rgb}{0.15,0.15,0.55}
	\definecolor{DarkPurple}{rgb}{0.4,0.2,0.4}

\ifnum\final=0
\newcommand{\mynote}[2]{{\color{#1} \marginpar{\tiny #2}}}
\newcommand{\mybignote}[2]{{\color{#1} $\langle \langle$ #2$\rangle \rangle$}}
\else
\newcommand{\mynote}[2]{}
\newcommand{\mybignote}[2]{}
\fi
\newcommand{\jnote}[1]{\mynote{DarkRed}{Jon: {#1}}}
\newcommand{\bigjnote}[1]{\mybignote{DarkRed}{Jon: #1}}
\newcommand{\mnote}[1]{\mynote{DarkBlue}{Mark: {#1}}}

\newcommand{\biglnote}[1]{\mybignote{DarkGreen}{Luke: #1}}
\newcommand{\tnote}[1]{\mynote{magenta}{Tal: {#1}}}

\newcommand{\INDSTATE}[1][1]{\STATE\hspace{#1\algorithmicindent}}

\newcommand{\pr}[2]{\underset{#1}{\mathbb{P}}\left[ #2 \right]}
\newcommand{\prob}[1]{\mathbb{P}\left[ #1 \right]}
\newcommand{\ex}[2]{\underset{#1}{\mathbb{E}}\left[ #2 \right]}

\newcommand{\poly}{\mathrm{poly}}

\newcommand{\zo}{\{0,1\}}
\newcommand{\bits}{\zo}

\newcommand{\getsr}{\gets_{\mbox{\tiny R}}}

\newcommand{\set}[1]{\left\{#1\right\}}
\newcommand{\from}{:}
\newcommand{\negl}{\mathrm{negl}}
\newcommand{\eps}{\varepsilon}

\newcommand{\ind}[1]{\mathbb{I}\{#1\}}

\newcommand{\N}{\mathbb{N}}

\newcommand{\cQ}{\mathcal{Q}}

\newcommand{\cS}{\mathcal{S}}

\ifnum\lncs=1
\else
\newtheorem{theorem}{Theorem}[section]

\newtheorem{lemma}[theorem]{Lemma}

\newtheorem{claim}[theorem]{Claim}

\theoremstyle{definition}

\newtheorem{definition}[theorem]{Definition}
\fi

\newcommand{\setup}{\ensuremath{\mathsf{Setup}}}
\newcommand{\prg}{\ensuremath{\mathsf{PRG}}}
\newcommand{\prf}{\ensuremath{\mathsf{PRF}}}

\newcommand{\sk}{\ensuremath{\mathit{sk}}}
\newcommand{\mk}{\ensuremath{\mathit{mk}}}
\newcommand{\s}{\ensuremath{s}}
\newcommand{\keys}{\ensuremath{\vec{k}}}
\newcommand{\enc}{\ensuremath{\mathsf{Enc}}}
\newcommand{\dec}{\ensuremath{\mathsf{Dec}}}
\newcommand{\obf}{\ensuremath{\mathsf{O}}}

\newcommand{\pdec}{\ensuremath{\mathsf{P}}}
\newcommand{\game}[1]{\ensuremath{\mathbf{#1}}}
\newcommand{\adv}{\ensuremath{\mathrm{Adv}}}
\newcommand{\twoadv}{\ensuremath{\mathrm{TwoAdv}}}
\newcommand{\indexhiding}{\game{IndexHiding}}
\newcommand{\twoindexhiding}{\game{TwoIndexHiding}}
\newcommand{\prfsetup}{\mathit{PRFSetup}}
\newcommand{\prfpuncture}{\mathsf{Puncture}}
\newcommand{\univ}{X}

\newcommand{\exponent}{7}
\newcommand{\otherexponent}{6}

\ifnum\lncs=0
	\title{Strong Hardness of Privacy from  Weak Traitor Tracing}
\author{Lucas Kowalczyk\thanks{Columbia University, Department of
Computer Science.\texttt{luke@cs.columbia.edu}.} \and Tal Malkin\thanks{Columbia University, Department of Computer Science. \texttt{tal@cs.columbia.edu}.} \and Jonathan Ullman\thanks{Northeastern University College of Computer and Information Science. \texttt{jullman@ccs.neu.edu}.} \and
Mark Zhandry\thanks{MIT EECS. \texttt{mzhandry@gmail.com}.}} 
\else
	\title{Strong Hardness of Privacy \\ from Weak Traitor Tracing\thanks{Full version.}}
	\author{Lucas Kowalczyk\inst{1} \and Tal Malkin\inst{1} \and Jonathan Ullman\inst{2} \and Mark Zhandry\inst{3}}
	\institute{Columbia University \and Northeastern University \and MIT}
\fi

\begin{document}

\maketitle

\ifnum\lncs=0
	\pagenumbering{gobble}
\else \fi
\begin{abstract}
Despite much study, the computational complexity of differential
privacy remains poorly understood.  In this paper we consider the
computational complexity of accurately answering a family $Q$
of \emph{statistical queries} over a \emph{data universe} $X$ under
differential privacy.  A statistical query on a dataset $D \in X^n$
asks ``what fraction of the elements of $D$ satisfy a given predicate
$p$ on $X$?''  Dwork et al. (STOC'09) and Boneh and Zhandry
(CRYPTO'14) showed that if both $Q$ and $X$ are of polynomial size,
then there is an efficient differentially private algorithm that
accurately answers all the queries, and if both $Q$ and $X$ are
exponential size, then under a plausible assumption, no efficient
algorithm exists. 

We show that, under the same assumption,
if \emph{either} the number of queries \emph{or} the data universe is of
exponential size, then there is no differentially private algorithm
that answers all the queries.
Specifically, we prove that if one-way functions and
indistinguishability obfuscation exist, then: 
\begin{enumerate}
\item For every $n$, there is a family $Q$ of $\tilde{O}(n^7)$ queries on a data universe $X$ of size $2^d$ such that no $\poly(n,d)$ time differentially private algorithm takes a dataset $D \in X^n$ and outputs accurate answers to every query in $Q$.

\item For every $n$, there is a family $Q$ of $2^d$ queries on a data universe $X$ of size $\tilde{O}(n^7)$ such that no $\poly(n,d)$ time differentially private algorithm takes a dataset $D \in X^n$ and outputs accurate answers to every query in $Q$.
\end{enumerate}

In both cases, the result is nearly quantitatively tight, since there
is an efficient differentially private algorithm that answers
$\tilde{\Omega}(n^2)$ queries on an exponential size data universe,
and one that answers exponentially many queries on a data universe of
size $\tilde{\Omega}(n^2)$. 

Our proofs build on the connection between hardness results in
differential privacy and traitor-tracing schemes (Dwork et al.,
STOC'09; Ullman, STOC'13).  We prove our hardness result for a
polynomial size query set (resp., data universe) by showing that they
follow from the existence of a special type of traitor-tracing scheme
with very short ciphertexts (resp., secret keys), but very weak
security guarantees, and then constructing such a scheme. 
\end{abstract}

\ifnum\lncs=0
\vfill
\newpage

\tableofcontents

\vfill
\newpage

\pagenumbering{arabic}
\else \fi
\section{Introduction}
The goal of privacy-preserving data analysis is to 
release rich statistical information about a sensitive dataset while
respecting the privacy of the individuals represented in that dataset.
The past decade has seen tremendous progress towards understanding
when and how these two competing goals can be reconciled, including
surprisingly powerful differentially private algorithms as well
as computational and information-theoretic limitations.  In this work,
we further this agenda by showing a strong new computational
bottleneck in differential privacy. 

Consider a dataset $D \in \univ^n$ where each of the $n$ elements is one individual's data, and each individual's data comes from some \emph{data universe} $X$.  We would like to be able to answer sets of \emph{statistical queries} on $D$, which are queries of the form ``What fraction of the individuals in $D$ satisfy some property $p$?''  However, \emph{differential privacy}~\cite{DworkMNS06} requires that we do so in such a way that no individual's data has significant influence on the answers.

If we are content answering a relatively small set of queries $Q$,
then it suffices to perturb the answer to each query with independent
noise from an appropriate distribution.  This algorithm is simple,
very efficient, differentially private, and ensures good
accuracy---say, within $\pm .01$
of the true answer---as long as $|Q| \lesssim n^2$
queries~\cite{DinurN03,DworkN04,BlumDMN05,DworkMNS06}. \tnote{Is this ``for large enough $n$''?}
\jnote{Yes, but isn't that what ``if $|Q| \lesssim n^2$'' means?  Is there a different thing you find confusing?}  \tnote{Hi Jon! The reason for my question was that I was surprised initially that the accuracy was a constant not dependent on $n$  -- I was imagining $n=1$ rows, and you can't answer accurately even a single query. But indeed, this is addressed by the dependency of $|Q|$ on $n$, so I have no issues (although $\lesssim$ is not something formal, I do find it clear)} 

Remarkably, the work of Blum, Ligett, and Roth~\cite{BlumLR08} showed
that it is possible 
to output a summary that allows accurate answers to 
an \emph{exponential} number of
queries---nearly $2^n$---while ensuring differential privacy.
However, neither their algorithm nor the subsequent
improvements~\cite{DworkNRRV09,DworkRV10,RothR10,HardtR10,GuptaRU12,NikolovTZ13,Ullman15}
are computationally efficient.  Specifically, they all require time 
at least $\poly(n, |\univ|, |Q|)$ to privately and accurately answer a family
of statistical queries $Q$ on a dataset $D \in \univ^{n}$.  Note that
the size of the input is $n \log|\univ|$ bits, so a computationally
efficient algorithm runs in time $\poly(n,\log |\univ|)$.\footnote{It
may require exponential time just to describe and evaluate an
arbitrary counting query, which would rule out efficiency for reasons
that have nothing to do with privacy.  In this work, we restrict
attention to queries that are efficiently computable in time
$\poly(n,\log|\univ|)$, so they are not the bottleneck in the
computation.}
For example, in the common setting where each individual's data
consists of $d$ binary attributes, namely $X = \zo^{d}$,  the size
of the input is $nd$ but $|\univ| = 2^d$. As a result, all known
private algorithms for answering arbitrary sets of statistical queries
are inefficient if either the number of queries or the size of the
data universe is superpolynomial.   

This accuracy vs.~computation tradeoff has been the subject of
extensive study.  Dwork et al.~\cite{DworkNRRV09} showed that the
existence of cryptographic \emph{traitor-tracing
schemes}~\cite{ChorFN94} yields a family of statistical queries that
cannot be answered accurately and efficiently with differential
privacy.  Applying recent traitor-tracing schemes~\cite{BonehZ14}, we
conclude that,
under plausible cryptographic assumptions (discussed below),
if both the
number of queries and the data universe can be superpolynomial, then
there is no efficient differentially private algorithm.
\cite{Ullman13} used variants of traitor-tracing schemes to show that
in the interactive setting, where the queries are not fixed but are
instead given as input to the algorithm, assuming one-way
functions 
exist, there is no private and efficient algorithm
that accurately answers more than $\tilde{O}(n^2)$ statistical
queries.  All of the algorithms mentioned above work in this
interactive setting, but for many applications we only need to answer a fixed family of statistical queries. 

Despite the substantial progress, there is still a basic gap in our
understanding.  The hardness results for Dwork et al.~apply
if \emph{both} the number of queries and the universe are large.  But
the known algorithms require exponential time if \emph{either} of
these sets is large.  Is this necessary?  Are there algorithms that
run in time $\poly(n, \log|\univ|, |Q|)$ or $\poly(n, |\univ|, \log
|Q|)$?   

Our main result shows that under the same plausible cryptographic
assumptions, the answer is no---if either the data universe or the set
of queries can be superpolynomially large, then there is some family of
statistical queries that cannot be accurately and efficiently answered
while ensuring differential privacy. 

\subsection{Our Results}
Our first result shows that if the data universe can be of
superpolynomial size then there is some fixed family of polynomially
many queries that cannot be efficiently answered under differential
privacy.  This result shows that the efficient algorithm for answering
an arbitrary family of $|Q| \lesssim n^2$ queries
by adding independent noise
is optimal up to the specific constant in the exponent. 
\begin{theorem}[Hardness for small query sets] \label{thm:mainshortctext}
Assume the existence of indistinguishability obfuscation and one-way
functions.  Let $\lambda \in \N$ be a computation parameter.  For any
polynomial $n = n(\lambda)$, there is a sequence of pairs $\{
(\univ_\lambda, Q_\lambda) \}$ with $|\univ_\lambda| = 2^\lambda$ and
$|Q_\lambda| = \tilde{O}(n^{\exponent})$ such that
there is 
no polynomial time differentially private algorithm
that 
takes a dataset $D \in \univ_{\lambda}^n$ and outputs an accurate
answer to every query in $Q_\lambda$ up to an additive error of $\pm
1/3$.  
\end{theorem}

Our second result shows that, even if the data universe is required to
be of polynomial size, there is a fixed set of superpolynomially
many queries that cannot be answered efficiently under differential
privacy.  When we say that an algorithm efficiently answers a set of
superpolynomially many queries, we mean that it efficiently outputs a
summary such that there is an efficient algorithm for obtaining an
accurate answer to any query in the set.
\tnote{ this comparison simple quadratic universe upper bound maybe
should have been in the preemble, parallel to the quadratic number of
queries simple upper bound?}  
For comparison, if $|\univ| \lesssim n^{2}$, then there is a simple
$\poly(n, |\univ|)$ time differentially private algorithm that
accurately answers superpolynomially many
queries.\ifnum\lncsshort=0\footnote{The algorithm, sometimes called
the \emph{noisy histogram algorithm}, works as follows. \tnote{reference?} 
First, convert the dataset $D$ to a vector
$(D_{x})_{x \in \univ}$ where $D_{x}$ is the fraction of $D$'s
elements that are equal to $x$.  Then, output a vector $\tilde{D} =
(\tilde{D}_{x})_{x \in \univ}$ where $\tilde{D}_x$ is equal to $D_{x}$
plus independent noise from an appropriately scaled Gaussian
distribution.  To answer a statistical query defined by a predicate
$p$, construct the vector $\tilde{p} = (p(x))_{x \in \univ}$ and
compute the answer $\langle \tilde{D}, \tilde{p}\rangle$.  One can
show that this algorithm is differentially private and for any fixed
set \tnote{why only fixed? or also interactive?} 
of statistical queries $Q$, with high probability, the maximum
error is $\tilde{O}(\sqrt{|\univ| \log |Q|} / n )$.  The running time
is $\poly(n, |\univ|)$ to construct $\tilde{D}$ and to evaluate each
query.} \else{\ } \fi
Our result shows that this efficient algorithm is optimal
up to the specific constant in the exponent. 
\begin{theorem}[Hardness for small query sets] \label{thm:mainshortkey}
Assume the existence of indistinguishability obfuscation and one-way
functions.  Let $\lambda \in \N$ be a computation parameter.  For any
polynomial $n = n(\lambda)$, there is a sequence of pairs $\{
(\univ_\lambda, Q_\lambda) \}$ with $|\univ_\lambda|
= \tilde{O}(n^{\exponent})$ and $|Q_\lambda| = 2^\lambda$ such that
there is 
no polynomial time differentially private algorithm
that 
takes a dataset $D \in \univ_{\lambda}^n$ and outputs an accurate
answer to every query in $Q_\lambda$ up to an additive error of $\pm
1/3$.  
\end{theorem}

Before we proceed to describe our techniques, we make a few remarks about these results.  In both of these results, the constant $1/3$ in our result is arbitrary, and can be replaced with any constant smaller than $1/2$.  We also remark that, when we informally say that an algorithm is differentially private, we mean that it satisfies $(\eps, \delta)$-differential privacy for some $\eps = O(1)$ and $\delta = o(1/n)$.  These are effectively the largest parameters for which differential privacy is a meaningful notion of privacy.  That our hardness results apply to these parameters only makes our results stronger.

\paragraph{On Indistinguishability Obfuscation.}
Indistinguishability obfuscation (iO) has recently become a central
cryptographic primitive. The first candidate construction, 
proposed just a couple years ago~\cite{GGHRSW13}, was followed by a
flurry of results demonstrating the extreme power and wide
applicability of iO (cf.,~\cite{GGHRSW13,SW14,BonehZ14,HSW14,BPW16}).
However, the assumption that iO exists is currently poorly understood,
and the debate over the plausibility of iO is far from settled.
While some specific proposed iO schemes have been
attacked~\cite{CGHLMMRST15,MSZ16}, other schemes seem to resist
all \emph{currently known} attacks~\cite{BMSZ16,GMS16}.
We also do not know how to base iO on a solid, simple, natural computational
assumption (some attempts based on multilinear maps have been
made~\cite{GLSW15}, but they were broken with respect 
to all current multilinear map constructions). 
\tnote{add citations here ~\cite{...}} 

Nevertheless, our results are meaningful whether or not iO exists.  If
iO exists, our results show that certain tasks in differential privacy
are intractable.  Interestingly, unlike many previous results relying
on iO, these conclusions were not previously known to follow from even
the much stronger (and in fact, false) assumption of virtual black-box
obfuscation.  If, on the other hand, iO does not exist, then our
results still demonstrate a barrier to progress in differential
privacy---such progress would need to \emph{prove} that iO does not
exist. Alternatively, our results highlight a possible path toward
proving that iO does not exist.  We note that other
``incompatibility'' results are known for iO; for example, iO and
certain types of hash functions cannot simultaneously
exist~\cite{BFM14,BST16}.

\subsection{Techniques}
We prove our results by building on the connection between
differentially private algorithms for answering statistical queries
and traitor-tracing schemes discovered by Dwork et
al.~\cite{DworkNRRV09}. Traitor-tracing schemes were introduced by
Chor, Fiat, and Naor~\cite{ChorFN94} for the purpose of identifying
pirates who violate copyright restrictions. Roughly speaking, a (fully
collusion-resilient) traitor-tracing scheme allows a sender to
generate keys for $n$ users so that 1) the sender can broadcast encrypted messages that can be decrypted by any user, and 2) any efficient pirate decoder capable of decrypting messages can be traced to at least one of the users who contributed a key to it, even if an arbitrary coalition of the users combined their keys in an arbitrary efficient manner to construct the decoder.  

Dwork et al.~show that the existence of traitor-tracing schemes implies hardness results for differential privacy. Very informally, they argue as follows.  Suppose a coalition of users takes their keys and builds a dataset $D \in X^n$ where each element of the dataset contains one of their user keys.  The family $Q$ will contain a query $q_c$ for each possible ciphertext $c$.  The query $q_c$ asks ``What fraction of the elements (user keys) in $D$ would decrypt the ciphertext $c$ to the message $1$?''  Every user can decrypt, so if the sender encrypts a message $b \in \zo$ as a ciphertext $c$, then every user will decrypt $c$ to $b$.  Thus, the answer to the statistical query $q_c$ will be $b$.  

Suppose there were an efficient algorithm that outputs an accurate answer to each query $q_c$ in $Q$.  Then the coalition could use it to efficiently produce a summary of the dataset $D$ that enables one to efficiently compute an approximate answer to every query $q_c$, which would also allow one to efficiently decrypt the ciphertext.  Such a summary can be viewed as an efficient pirate decoder, and thus the tracing algorithm can use the summary to trace one of the users in the coalition.  However, if there is a way to identify one of the users in the dataset from the summary, then the summary is not differentially private.

To instantiate this result, they need a traitor-tracing scheme.  Observe that the data universe contains one element for every possible user key, and the set of queries contains one query for every ciphertext, and we want to minimize the size of these sets.  Boneh and Zhandry constructed a traitor-tracing scheme where both the keys and the ciphertexts have length equal to the security parameter $\lambda$, which yields hardness for a data universe and query set each of size $2^{\lambda}$.  The main contribution of this work is to show that we can reduce either the number of possible ciphertexts or the number of possible keys to $\poly(n)$ while the other remains of size $2^{\lambda}$.

Suppose we want to reduce the number of possible ciphertexts to
$\poly(n)$.  How can we possibly have a secure traitor-tracing scheme
with only polynomially many ciphertexts, when even a semantically
secure private key encryption scheme requires superpolynomially many
ciphertexts?  The answer lies in an observation from~\cite{Ullman13}
that in order to show hardness for differential privacy, it suffices
to have a traitor-tracing scheme with extremely weak security.  First,
in the reduction from differential privacy to breaking traitor-tracing
the adversary has to produce the pirate decoder using only the
coalition's user keys and does not have access to an encryption
oracle.  Second, the probability that tracing fails only needs to be
$o(1/n)$, rather than negligible.  Both of these relaxations of the
standard definition of traitor-tracing are crucial to making the
ciphertext size $\poly(n)$, and as we show, these two relaxations are
in fact sufficient.  Alternatively, we can use these relaxations also
allow us to reduce the key size to $\poly(n)$.
\ifnum\lncsshort=1
We defer the reader to Sections~\ref{sec:shortctext}
and~\ref{sec:shortkey}
\else 
We defer the reader to the constructions of Sections~\ref{sec:shortctext}
and~\ref{sec:shortkey}
\fi 
for more details about how we achieve this goal. 

\ifnum\lncsshort=0
\subsection{Related Work}
\tnote{Here (long version) I'd add something about Dinur-nissim attacks}
Theorem~\ref{thm:mainshortctext} should be contrasted with the line of work on answering \emph{width-$w$ marginal queries} under differential privacy~\cite{GuptaHRU11,HardtRS12,ThalerUV12,ChandrasekaranTUW14,DworkNT14}.  A width-$w$ marginal query is defined on the data universe $\zo^{\lambda}$.  It is specified by a set of positions $S \subseteq \set{1,\dots,\lambda}$ of size $w$, and a pattern $t \in \zo^{w}$ and asks ``What fraction of elements of the dataset have each coordinate $j \in S$ set to $t_j$?''  Specifically, Thaler, Ullman, and Vadhan~\cite{ThalerUV12}, building on the work of Hardt, Rothblum, and Servedio~\cite{HardtRS12} gave an efficient differentially private algorithm for answering $n^{\Omega(\sqrt{w})} \gg n^{\exponent}$ width-$w$ marginal queries up to an additive error of $\pm .01$.  There are also computationally efficient algorithms that answer exponentially many queries from even simpler families like \emph{point queries} and \emph{threshold queries}~\cite{BeimelNS13,BunNSV15}.  

There have been several other attempts to explain the accuracy
vs.~computation tradeoff in differential privacy by considering
restricted classes of algorithms.  For example, Ullman and
Vadhan~\cite{UllmanV11} (building on Dwork et al.~\cite{DworkNRRV09})
show that, assuming one-way functions, no differentially private and
computationally efficient algorithm that outputs a \emph{synthetic
dataset} can accurately answer even the very simple family of $2$-way
marginals.  \tnote{explain 2-way marginals and synthetic dataset?} 
This result is incomparable to ours, since it applies to a very small and simple family of statistical queries, but necessarily only applies to algorithms that output synthetic data.

Gupta et al.~\cite{GuptaHRU11} showed that no algorithm can obtain accurate answers to all marginal queries just by asking a polynomial number of statistical queries on the dataset.  Thus, any algorithm that can be implemented using only statistical queries, even one that is not differentially private, can run in polynomial time.

\ifnum\lncsshort=0
Bun and Zhandry considered the incomparable problem
of \emph{differentially private PAC learning}~\cite{BunZ16} and showed
that there is a concept class that is efficiently PAC learnable and
inefficiently PAC learnable under differential privacy, but is not
efficiently PAC learnable under differential privacy, settling an open question of Kasvisiwanathan et al.~\cite{KasiviswanathanLNRS09}, who introduced the model of differentially private PAC learning.
\fi

There is also a line of work using \emph{fingerprinting codes} to
prove \emph{information-theoretic} lower bounds on differentially
private mechanisms~\cite{BunUV14,SteinkeU15b,DworkSSUV15}.  Namely,
that if the data universe is of size $\exp(n^2)$, then there is no
differentially private algorithm, even a computationally unbounded
one, that can answer more than $n^2$ statistical queries.
Fingerprinting codes are essentially the information-theoretic
analogue of traitor-tracing schemes, and thus these results are
technically related, although the models are incomparable. 

Finally, we remark that techniques for proving hardness results in differential privacy have also found applications to the problem of \emph{interactive data analysis}~\cite{HardtU14,SteinkeU15a}.  The technical core of these results is to show that if an adversary is allowed to ask an online sequence of adaptively chosen statistical queries, then he can not only recover one element of the dataset, but can actually recover every element of the dataset.  Doing so rules out any reasonable notion of privacy, and makes many non-private learning tasks impossible.  The results are proven using variants of the sorts of traitor-tracing schemes that we study in this work.
\else

\subsection{Related Work}
Theorem~\ref{thm:mainshortctext} should be contrasted with the line of
work showing that differentially private algorithms can efficiently
answer many more than $n^2$ \emph{simple} queries.  These results
include algorithms for highly structured queries like point queries,
threshold queries, and conjunctions (see
e.g.~\cite{ThalerUV12,BeimelNS13} and the references therein). 

Ullman and Vadhan~\cite{UllmanV11} (building on Dwork et al.~\cite{DworkNRRV09}) show that, assuming one-way functions, no differentially private and computationally efficient algorithm that outputs a \emph{synthetic dataset} can accurately answer even the very simple family of $2$-way marginals.  This result is incomparable to ours, since it applies to a very small and simple family of statistical queries, but necessarily only applies to algorithms that output synthetic data.

There is also a line of work using \emph{fingerprinting codes} to
prove \emph{information-theoretic} lower bounds on differentially
private mechanisms~\cite{BunUV14,SteinkeU15b,DworkSSUV15}.  Namely,
that if the data universe is of size $\exp(n^2)$, then there is no
differentially private algorithm, even a computationally unbounded
one, that can answer more than $n^2$ statistical queries.
Fingerprinting codes are essentially the information-theoretic
analogue of traitor-tracing schemes, and thus these results are
technically related, although the models are incomparable. 
\fi

\ifnum\lncsshort=1
\subsection{Paper Outline}
In Section~\ref{sec:dpprelims} we will give the necessary background on differential privacy.  In Section~\ref{sec:ttschemes} we will give our definition of traitor-tracing schemes and in Section~\ref{sec:tttodp} we will connect traitor-tracing schemes to differential privacy.  In Section~\ref{sec:cryptoprimitives} we will define some cryptographic tools that we use to construct traitor tracing schemes.  In Section~\ref{sec:shortctext} we will construct the short-ciphertext scheme we use to prove Theorem~\ref{thm:mainshortctext}.  Due to space, we omit the full construction of the short-key scheme that we use to prove Theorem~\ref{thm:mainshortkey}, but in Section~\ref{sec:shortkey} we briefly state the main ideas.
\fi

\section{Differential Privacy Preliminaries} \label{sec:dpprelims}
\subsection{Differentially Private Algorithms}
A \emph{dataset} $D \in \univ^{n}$ is an ordered set of $n$ rows, where each row corresponds to an individual, and each row is an element of some the \emph{data universe} $\univ$.  We write $D = (D_1,\dots,D_n)$ where $D_i$ is the $i$-th row of $D$.  We will refer to $n$ as the \emph{size} of the dataset.
We say that two datasets $D, D' \in \univ^*$ are \emph{adjacent} if $D'$ can be obtained from $D$ by the addition, removal, or substitution of a single row, and we denote this relation by $D \sim D'$.  In particular, if we remove the $i$-th row of $D$ then we obtain a new dataset $D_{-i} \sim D$.
Informally, an algorithm $A$ is differentially private if it is randomized and for any two adjacent datasets $D \sim D'$, the distributions of $A(D)$ and $A(D')$ are similar.
\begin{definition}[Differential Privacy~\cite{DworkMNS06}]\label{def:dp} 
Let $A \from \univ^n \to S$ be a randomized algorithm.  We say that $A$ is \emph{$(\eps, \delta)$-differentially private} if for every two adjacent datasets $D \sim D'$ and every subset $T \subseteq S$,
\ifnum\lncsshort=1$\else$$\fi
\pr{}{A(D) \in T} \leq e^{\eps} \cdot \pr{}{A(D') \in T} + \delta.
\ifnum\lncsshort=1$\else$$\fi\linebreak
In this definition, $\eps, \delta$ may be a function of $n$.
\end{definition}

\subsection{Algorithms for Answering Statistical Queries}
In this work we study algorithms that answer \emph{statistical queries} (which are also sometimes called \emph{counting queries}, \emph{predicate queries}, or \emph{linear queries} in the literature).  For a data universe $\univ$, a statistical query on $\univ$ is defined by a predicate $q \from \univ \to \zo$.  Abusing notation, we define the evaluation of a query $q$ on a dataset $D = (D_1,\dots,D_n) \in \univ^n$ to be
\ifnum\lncsshort=1$\else$$\fi
\frac{1}{n} \sum_{i=1}^{n} q(D_i).
\ifnum\lncsshort=1$\else$$\fi

A single statistical query does not provide much useful information about the dataset.  However, a sufficiently large and rich set of statistical queries is sufficient to implement many natural machine learning and data mining algorithms~\cite{Kearns93}, thus we are interesting in differentially private algorithms to answer such sets.  To this end, let $Q = \set{q \from \univ \to \zo}$ be a set of statistical queries on a data universe $\univ$.   

Informally, we say that a mechanism is accurate for a set $Q$ of statistical queries if it answers every query in the family to within error $\pm \alpha$ for some suitable choice of $\alpha > 0$.  Note that $0 \leq q(D) \leq 1$, so this definition of accuracy is meaningful when $\alpha < 1/2$.

Before we define accuracy, we note that the mechanism may represent its answer in any form.  That is, the mechanism outputs may output a \emph{summary} $S \in \cS$ that somehow represents the answers to every query in $Q$.  We then require that there is an \emph{evaluator} $\mathit{Eval} \from \cS \times \cQ \to [0,1]$ that takes the summary and a query and outputs an approximate answer to that query.  That is, we think of $\mathit{Eval}(S,q)$ as the mechanism's answer to the query $q$.  We will abuse notation and simply write $q(S)$ to mean $\mathit{Eval}(S,q)$.\footnote{If we do not restrict the running time of the algorithm, then it is without loss of generality for the algorithm to simply output a list of real-valued answers to each queries by computing $\mathit{Eval}(S,q)$ for every $q \in Q$.  However, this transformation makes the running time of the algorithm at least $|Q|$.  The additional generality of this framework allows the algorithm to run in time sublinear in $|Q|$.  Using this framework is crucial, since some of our results concern settings where the number of queries is exponential in the size of the dataset.}
\begin{definition} [Accuracy]
For a family $Q$ of statistical queries on $\univ$, a dataset $D \in \univ^n$ and a summary $s \in S$, we say that \emph{$s$ is $\alpha$-accurate for $Q$ on $D$} if
\ifnum\lncsshort=1
for every $q \in Q$, $\left| q(D) - q(s) \right| \leq \alpha.$
\else
$$
\forall q \in Q ~~~~ \left| q(D) - q(s) \right| \leq \alpha.
$$
\fi
For a family of statistical queries $Q$ on $\univ$, we say that an algorithm $A \from \univ^n \to S$ is \emph{$(\alpha, \beta)$-accurate for $Q$ given a dataset of size $n$} if for every $D \in \univ^n$,
\ifnum\lncsshort=1$\else$$\fi
\pr{}{\textrm{$A(D)$ is $\alpha$-accurate for $Q$ on $\univ$}} \geq 1-\beta.
\ifnum\lncsshort=1$\else$$\fi
\end{definition}

In this work we are typically interested in mechanisms that satisfy the very weak notion of $(1/3, o(1/n))$-accuracy, where the constant $1/3$ could be replaced with any constant $< 1/2$.  Most differentially private mechanisms satisfy quantitatively much stronger accuracy guarantees.  Since we are proving hardness results, this choice of parameters makes our results stronger.

\subsection{Computational Efficiency}
Since we are interested in asymptotic efficiency, we introduce a computation parameter $\lambda \in \N$.  We then consider a sequence of pairs $\{ (X_{\lambda}, Q_{\lambda}) \}_{\lambda \in \N}$ where $Q_{\lambda}$ is a set of statistical queries on $X_{\lambda}$.  We consider databases of size $n$ where $n = n(\lambda)$ is a polynomial.  We then consider algorithms $A$ that take as input a dataset $X_{\lambda}^n$ and output a summary in $S_{\lambda}$ where $\{ S_{\lambda} \}_{\lambda \in \N}$ is a sequence of output ranges.  There is an associated evaluator $\mathit{Eval}$ that takes a query $q \in Q_{\lambda}$ and a summary $s \in S_{\lambda}$ and outputs a real-valued answer.  The definitions of differential privacy and accuracy extend straightforwardly to such sequences.

We say that such an algorithm is \emph{computationally efficient} if the running time of the algorithm and the associated evaluator run in time polynomial in the computation parameter $\lambda$.
\ifnum\lncsshort=0
\footnote{The constraint that the evaluator run in polynomial time sounds academic, but is surprisingly crucial.  For any $Q$ on $X$, there is an extremely simple differentially private algorithm that runs in time $\poly(n,|Q|)$ and outputs a summary that is accurate for $Q$, yet the summary takes time $\poly(|X|,|Q|)$ to evaluate~\cite{NikolovTZ13}.} \fi We remark that in principle, it could require at many as $|X|$ bits even to specify a statistical query, in which case we cannot hope to answer the query efficiently, even ignoring privacy constraints.  In this work we restrict attention exclusively to statistical queries that are specified by a circuit of size $\poly(\log |X|)$, and thus can be evaluated in time $\poly(\log |X|)$, and so are not the bottleneck in computation.  To remind the reader of this fact, we will often say that $\cQ$ is a family of \emph{efficiently computable statistical queries}.

\ifnum\lncsshort=0
\subsection{Notational Conventions}
Given a boolean predicate $p$, we will write $\ind{p}$ to denote the value $1$ if $p$ is true and $0$ if $p$ is false.  Also, given a vector $\vec{v} = (v_1,\dots,v_n) \in X^n$ and an index $i \in [n]$, we will use $v_{-i}$ to denote the vector $\vec{v}_{-i} = (v_1,\dots,v_{i-1},\bot, v_{i+1},\dots,v_{n}) \in X^{n}$ in which the $i$-th element of $\vec{v}$ is replaced by some unspecified fixed element of $X$ denoted $\bot$.  We also say that a function $f$ is \emph{negligible}, and write $f(n) = \negl(n)$, if $f(n) = O(1/n^c)$ for every constant $c > 0$.
\fi

\section{Weakly Secure Traitor-Tracing Schemes} \label{sec:ttschemes}
In this section we describe a very relaxed notion of traitor-tracing schemes whose existence will imply the hardness of differentially private data release.

\subsection{Syntax and Correctness}
For a function $n \from \N \to \N$ and a sequence $\{ K_\lambda, C_\lambda \}_{\lambda \in \N}$, an \emph{$(n, \{K_\lambda, C_\lambda \})$-traitor-tracing scheme} is a tuple of efficient algorithms $\Pi = (\setup, \enc, \dec)$ with the following syntax.
\begin{itemize}
\item $\setup$ takes as input a security parameter $\lambda$, runs in time $\poly(\lambda)$, and outputs $n = n(\lambda)$ secret \emph{user keys} $\sk_1,\dots,\sk_n \in K_\lambda$ and a secret \emph{master key} $\mk$.  We will write $\keys = (\sk_1,\dots,\sk_n, \mk)$ to denote the set of keys.

\item $\enc$ takes as input a master key $\mk$ and an \emph{index} $i \in \set{0,1,\dots,n}$, and outputs a ciphertext $c \in C_\lambda$.  If $c \getsr \enc(j, \mk)$ then we say that $c$ is \emph{encrypted to index $j$.}

\item $\dec$ takes as input a ciphertext $c$ and a user key $\sk_i$ and outputs a single bit $b \in \zo$.  We assume for simplicity that $\dec$ is deterministic.
\end{itemize}

Correctness of the scheme asserts that if $\keys$ are generated by $\setup$, then for any pair $i,j$,  $\dec(\sk_i, \enc(\mk, j)) = \ind{i \leq j}$.  For simplicity, we require that this property holds with probability $1$ over the coins of $\setup$ and  $\enc$, although it would not affect our results substantively if we required only correctness with high probability.  
\begin{definition}[Perfect Correctness] \label{def:correctness}
An $(n, \{K_\lambda, C_\lambda\})$-traitor-tracing scheme is \emph{perfectly correct} if for every $\lambda \in \N$, and every $i,j \in \set{0,1,\dots,n}$
$$
\pr{\keys = \setup(\lambda), \, c = \enc(\mk, j)}{\dec(\sk_i, c) = \ind{i \leq j}} = 1.
$$
\end{definition}

\subsection{Index-Hiding Security}
Intuitively, the security property we want is that any computationally efficient adversary who is missing one of the user keys $\sk_{i^*}$ cannot distinguish ciphertexts encrypted with index $i^*$ from index $i^*-1$, even if that adversary holds all $n-1$ other keys $\sk_{-i^*}$.  In other words, an efficient adversary cannot infer anything about the encrypted index beyond what is implied by the correctness of decryption and the set of keys he holds.  

More precisely, consider the following two-phase experiment.  First the adversary is given every key except for $\sk_{i^*}$, and outputs a decryption program $S$.  Then, a challenge ciphertext is encrypted to either $i^*$ or to $i^*-1$.  We say that the traitor-tracing scheme is secure if for every polynomial time adversary, with high probability over the setup and the decryption program chosen by the adversary, the decryption program has small advantage in distinguishing the two possible indices.

\begin{definition}[Index Hiding] \label{def:indexhiding}
A traitor-tracing scheme $\Pi$ satisfies \emph{(weak) index-hiding security} if for every sufficiently large $\lambda \in \N$, every $i^* \in [n(\lambda)],$ and every adversary $A$ with running time $\poly(\lambda)$,
\begin{equation}\label{eq:indexhidingdef}
\pr{\keys = \setup(\lambda), \, S = A(\sk_{-i^*})}{ \pr{}{S(\enc(\mk, i^*)) = 1} - \pr{}{S(\enc(\mk, i^* - 1)) = 1} > \frac{1}{2en}} \leq \frac{1}{2en}
\end{equation}
In the above, the inner probabilities are taken over the coins of $\enc$ and $S$.
\end{definition}
Note that in the above definition we have fixed the success probability of the adversary for simplicity.  Moreover, we have fixed these probabilities to relatively large ones.  Requiring only a polynomially small advantage is crucial to achieving the key and ciphertext lengths we need to obtain our results, while still being sufficient to establish the hardness of differential privacy.

\subsubsection{The Index-Hiding and Two-Index-Hiding Games}
While Definition~\ref{def:indexhiding} is the most natural, in this section we consider some related ways of defining security that will be easier to work with when we construct and analyze our schemes.  Consider the following $\game{IndexHiding}$ game.
\begin{figure}[ht]
\begin{framed}
\begin{algorithmic}
\INDSTATE[0]{The challenger generates keys $\keys = (\sk_1,\dots,\sk_n,\mk) \getsr \setup(\lambda)$.}
\INDSTATE[0]{The adversary $A$ is given keys $\sk_{-i^*}$ and outputs a decryption program $S$.}
\INDSTATE[0]{The challenger chooses a bit $b \getsr \zo$}
\INDSTATE[0]{The challenger generates an encryption to index $i^*-  b$, $c \getsr \enc(\mk, i^* - b)$}
\INDSTATE[0]{The adversary makes a guess $b' = S(c)$}
\end{algorithmic}
\end{framed}
\vspace{-6mm}
\caption{$\game{IndexHiding}[i^*]$}
\end{figure}

Let $\indexhiding[i^*, \keys, S]$ be the game $\indexhiding[i^*]$ where we fix the choices of $\keys$ and $S$.  Also, define
$$
\adv[i^*, \keys, S] = \pr{\mathsf{IndexHiding}[i^*, \keys, S]}{b' = b} - \frac{1}{2}.
$$
so that
$$
\pr{\mathsf{IndexHiding}[i^*]}{b' = b} - \frac{1}{2} = \ex{\keys = \setup(\lambda) \atop S = A(\sk_{-i^*})}{\adv[i^*, \keys, S]}
$$
Then the following is equivalent to~\eqref{eq:indexhidingdef} in Definition~\ref{def:indexhiding} as
\begin{equation} \label{eq:indexhidingadv}
\pr{\keys = \setup(\lambda), \, S = A(\sk_{-i^*})}{\adv[i^*, \keys, S] > \frac{1}{4en}} \leq \frac{1}{2en}
\end{equation}

In order to prove that our schemes satisfy weak index-hiding security, we will go through an intermediate notion that we call two-index-hiding security.  To see why this is useful, In our constructions it will be fairly easy to prove that $\adv[i^*]$ is small, but because $\adv[i^*, \keys, S]$ can be positive or negative, that alone is not enough to establish~\eqref{eq:indexhidingadv}.  Thus, in order to establish~\eqref{eq:indexhidingadv} we will analyze the following variant of the index-hiding game.
\begin{figure}[ht]
\begin{framed}
\begin{algorithmic}
\INDSTATE[0]{The challenger generates keys $\keys = (\sk_1,\dots,\sk_n,\mk) \getsr \setup$.}
\INDSTATE[0]{The adversary $A$ is given keys $\sk_{-i^*}$ and outputs a decryption program $S$.}
\INDSTATE[0]{Choose $b_0 \getsr \bits$ and $b_1 \getsr \bits$ independently.}
\INDSTATE[0]{Let $c_0 \getsr \enc(i^* - b_0; \mk)$ and $c_1 \getsr \enc(i^* - b_1; \mk).$}
\INDSTATE[0]{Let $b' = S(c_0, c_1).$}
\end{algorithmic}
\end{framed}
\vspace{-6mm}
\caption{$\twoindexhiding[i^*]$}
\end{figure}

Analogous to what we did with \indexhiding, we can define $\twoindexhiding[i^*, \keys, S]$ to be the game $\twoindexhiding[i^*]$ where we fix the choices of $\keys$ and $S$, and define
\begin{align*}
\twoadv[i^*] ={} &\pr{\twoindexhiding[i^*]}{b' = b_0 \oplus b_1} - \frac{1}{2} \\
\twoadv[i^*, \keys, S] ={} &\pr{\twoindexhiding[i^*, \keys, S]}{b' = b_0 \oplus b_1} - \frac{1}{2}
\end{align*}
so that
$$
\pr{\twoindexhiding[i^*]}{b' = b_0 \oplus b_1} - \frac{1}{2} = \ex{\keys = \setup(\lambda), S = A(\sk_{-i^*})}{\twoadv[i^*, \keys, S]}
$$
The crucial feature is that if we can bound the expectation of $\twoadv$ then we get a bound on the expectation of $\adv^2$.  Since $\adv^2$ is always positive, we can apply Markov's inequality to establish~\eqref{eq:indexhidingadv}.  Formally, we have the following claim.
\begin{claim} \label{clm:2indextoindex}
Suppose that for every efficient adversary $A$, $\lambda \in \N$, and index $i^* \in [n(\lambda)],$
\ifnum\lncsshort=1$\else$$\fi
\twoadv[i^*] \leq \eps.
\ifnum\lncsshort=1$\else$$\fi
Then for every efficient adversary $A$, $\lambda \in \N$, and index $i^* \in [n(\lambda)],$
\begin{equation} \label{eq:boundedexadv1squared}
\ex{\keys = \setup(\lambda), S \gets A(\sk_{-i^*})}{\adv[i^*, \keys, S]^2} \leq \frac{\eps}{2}.
\end{equation}
\end{claim}

Using this claim we can prove the following lemma.
\begin{lemma} \label{lem:2indextoindex}
Let $\Pi$ be a traitor-tracing scheme such that for every efficient adversary $A$, $\lambda \in \N$, and index $i^* \in [n(\lambda)],$
\ifnum\lncsshort=1$\else$$\fi
\twoadv[i^*] \leq \frac{1}{200 n^3}.
\ifnum\lncsshort=1$\else$$\fi
Then $\Pi$ satisfies weak index-hiding security.
\end{lemma}
\ifnum\lncsshort=1
The proof is a simple calculation, which we omit for space.
\else
\begin{proof}
By applying Claim~\ref{clm:2indextoindex} to the assumption of the lemma, we have that for every efficient adversary $A$,
$$
\ex{\keys = \setup(\lambda), S = A(\sk_{-i^*})}{\adv[i^*, \keys, S]^2} \leq \frac{1}{400n^3}
$$
Now we have
\begin{align*}
&\ex{\keys = \setup(\lambda), S = A(\sk_{-i^*})}{\adv[i^*, \keys, S]^2} \leq \frac{1}{400n^3} \\
\Longrightarrow{} &\pr{\keys = \setup(\lambda), S = A(\sk_{-i^*})}{\adv[i^*, \keys, S]^2 > \frac{1}{(4en)^2}} \leq \frac{(4en)^2}{400n^3} \leq \frac{1}{2en} \tag{Markov's Inequality} \\
\Longrightarrow{} &\pr{\keys = \setup(\lambda), S = A(\sk_{-i^*})}{\adv[i^*, \keys, S] > \frac{1}{4en}} \leq \frac{\eps}{2(4en)^2}
\end{align*}
To complete the proof, observe that this final condition is equivalent to the definition of weak index-hiding security (Definition~\ref{def:indexhiding}).
\end{proof}

\fi
In light of this lemma, we will focus on proving that the schemes we construct in the following sections satisfying the condition 
\ifnum\lncsshort=1$\else$$\fi
\twoadv[i^*] \leq \frac{1}{200 n^3},
\ifnum\lncsshort=1$\else$$\fi
which will be easier than directly establishing Definition~\ref{def:indexhiding}.


\section{Hardness of Differential Privacy from Traitor Tracing} \label{sec:tttodp}
In this section we prove that traitor-tracing scheme satisfying perfect correctness and index-hiding security yields a family of statistical queries that cannot be answered accurately by an efficient differentially private algorithm.  The proof is a fairly straightforward adaptation of the proofs in Dwork et al.~\cite{DworkNRRV09} and Ullman~\cite{Ullman13} that various sorts of traitor-tracing schemes imply hardness results for differential privacy.  We include the result for completeness, and to verify that our very weak definition of traitor-tracing is sufficient to prove hardness of differential privacy.
\begin{theorem} \label{thm:tracingtodp}
Suppose there is an $(n, \{K_\lambda, C_\lambda\})$-traitor-tracing scheme that satisfies perfect correctness (Definition~\ref{def:correctness}) and index-hiding security (Definition~\ref{def:indexhiding}).  Then there is a sequence of of pairs $\{X_\lambda, Q_\lambda\}_{\lambda \in \N}$ where $Q_\lambda$ is a set of statistical queries on $X_\lambda$, $|Q_\lambda| = |C_\lambda|$, and $|X_\lambda| = |K_\lambda|$ such that there is no algorithm $A$ that is simultaneously,
\begin{enumerate}
\item $(1, 1/2n)$-differentially private,
\item $(1/3, 1/2n)$-accurate for $Q_\lambda$ on datasets $D \in X_{\lambda}^{n(\lambda)}$, and
\item computationally efficient.
\end{enumerate}
\end{theorem}
Theorem~\ref{thm:mainshortctext} and~\ref{thm:mainshortkey} in the
introduction follow by combining Theorem~\ref{thm:tracingtodp} above
with the constructions of traitor-tracing schemes in
\ifnum\lncsshort=1
Section~\ref{sec:shortctext}. 
\else
Sections~\ref{sec:shortctext} and~\ref{sec:shortkey}.
\fi
The proof of
Theorem~\ref{thm:tracingtodp} closely follows the proofs in Dwork et
al.~\cite{DworkNRRV09} and Ullman~\cite{Ullman13}.  We give the proof
both for completeness and to verify that our definition of
traitor-tracing suffices to establish the hardness of differential
privacy. 
\begin{proof}
Let $\Pi = (\setup, \enc, \dec)$ be the promised $(n, \{K_\lambda, C_\lambda\})$ traitor-tracing scheme.  For every $\lambda \in \N$, we can define a distribution on datasets $D \in X_{\lambda}^{n(\lambda)}$ as follows.  Run $\setup(\lambda)$ to obtain $n = n(\lambda)$ secret user keys $\sk_1,\dots,\sk_n \in K_\lambda$ and a master secret key $\mk$.  Let the dataset be $D = (\sk_1,\dots,\sk_n) \in X_{\lambda}^{n}$ where we define the data universe $X_{\lambda} = K_{\lambda}$.  Abusing notation, we'll write $(D, \mk) \getsr \setup(\lambda)$.

Now we define the family of queries $Q_\lambda$ on $X_\lambda$ as follows.  For every ciphertext $c \in C_\lambda$, we define the predicate $q_{c} \in Q_{\lambda}$ to take as input a user key $\sk_i \in K_{\lambda}$ and output $\dec(\sk_i, c)$.  That is,
\ifnum\lncsshort=1$\else$$\fi
Q_{\lambda} = \set{q_{c}(\sk) = \dec(\sk, c) \; \mid \; c \in C_\lambda}.
\ifnum\lncsshort=1$\else$$\fi
~Recall that, by the definition of a statistical query, for a dataset $D = (\sk_1,\dots,\sk_n)$, we have $$q_{c}(D) = (1/n) \sum_{i=1}^{n} \dec(\sk_i, c).$$

Suppose there is an algorithm $A$ that is computationally efficient and is $(1/3, 1/2n)$-accurate for $Q_{\lambda}$ given a dataset $D \in X_{\lambda}^{n}$.  We will show that $A$ cannot satisfy $(1, 1/2n)$-differential privacy.  By accuracy, for every $\lambda \in \N$ and every fixed dataset $D \in X_{\lambda}^{n}$, with probability at least $1-1/2n$, $A(D)$ outputs a summary $S \in \cS_{\lambda}$ that is $1/3$-accurate for $Q_{\lambda}$ on $D$.  That is, for every $D \in X_{\lambda}^{n}$, with probability at least $1-1/2n$,
\begin{equation} \label{eq:acc1}
\forall q_{c} \in Q_\lambda ~~~ \left| q_{c}(D) - q_{c}(S) \right| \leq 1/3.
\end{equation}
Suppose that $S$ is indeed $1/3$-accurate.  By perfect correctness of the traitor-tracing scheme (Definition~\ref{def:correctness}), and the definition of $Q$, we have that since $(D,\mk) = \setup(\lambda)$,
\begin{equation} \label{eq:acc2}
(c = \enc(\mk, 0)) \Longrightarrow (q_{c}(D) = 0) ~~~~~~~~~ (c = \enc(\mk, n)) \Longrightarrow (q_{c}(D) = 1).
\end{equation}
Combining Equations~\eqref{eq:acc1} and~\eqref{eq:acc2}, we have that if $(D, \mk) = \setup(\lambda)$, $S \getsr A(D)$, and $S$ is $1/3$-accurate, then we have both
\ifnum\lncsshort=1
$\pr{c \getsr \enc(\mk, 0)}{q_{c}(S) \leq 1/3} = 1$ and $\pr{c \getsr \enc(\mk, n)}{q_{c}(S) \leq 1/3} = 0$
\else
$$
\pr{c \getsr \enc(\mk, 0)}{q_{c}(S) \leq 1/3} = 1 ~~~~~~~~~~~~ \pr{c \getsr \enc(\mk, n)}{q_{c}(S) \leq 1/3} = 0
$$
\fi
Thus, for every $(D,\mk)$ and $S$ that is $1/3$-accurate, there exists an index $i \in \set{1,\dots,n}$ such that
\begin{equation} \label{eq:acc3}
\left| \pr{c \getsr \enc(\mk, i)}{q_{c}(S) \leq 1/3} - \pr{c \getsr \enc(\mk, i-1)}{q_{c}(S) \leq 1/3} \right| > \frac{1}{n}
\end{equation}
By averaging, using the fact that $S$ is $1/3$-accurate with probability at least $1-1/2n$, there must exist an index $i^* \in \set{1,\dots,n}$ such that
\begin{equation} \label{eq:acc4}
\pr{(D,\mk) = \setup(\lambda), S \getsr A(D)}{\left| \pr{c \getsr \enc(\mk, i^*)}{q_{c}(S) \leq 1/3} - \pr{c \getsr \enc(\mk, i^*-1)}{q_{c}(S) \leq 1/3} \right| > \frac{1}{n}} \geq \frac{1}{n}
\end{equation}

Assume, for the sake of contradiction that $A$ is $(1, 1/2n)$-differentially private.  For a given $i, \mk$, let $\cS_{i, \mk} \subseteq \cS_{\lambda}$ be the set of summaries $S$ such that~\eqref{eq:acc3} holds.  Then, by~\eqref{eq:acc4}, we have
$$
\pr{(D,\mk) \getsr \setup(\lambda)}{A(D) \in \cS_{i^*, \mk}} \geq \frac{1}{n}.
$$
By differential privacy of $A$, we have
$$
\pr{(D,\mk) \getsr \setup}{A(D_{-i^*}) \in \cS_{i^*, \mk}} \geq \frac{1}{e}\left(\frac{1}{n} - \frac{1}{2n}\right) = \frac{1}{2en} 
$$
Thus, by our definition of $\cS_{i^*, \mk}$, and by averaging over $(D, \mk) \getsr \setup(\lambda)$, we have
\begin{equation} \label{eq:acc5}
\pr{(D, \mk) = \setup, S \getsr A(D_{-i^*})}{\left| \pr{c \getsr \enc(\mk, i^*)}{q_{c}(S) \leq 1/3} - \pr{c \getsr \enc(\mk, i^*-1)}{q_{c}(S) \leq 1/3} \right| > \frac{1}{n}} \geq \frac{1}{2en}
\end{equation}
But this violates the index hiding property of the traitor tracing scheme.  Specifically, if we consider an adversary for the traitor tracing scheme that runs $A$ on the keys $\sk_{-i^*}$ to obtain a summary $S$, then decrypts a ciphertext $c$ by computing $q_{c}(S)$ and rounding the answer to $\zo$, then by~\eqref{eq:acc5} this adversary violates index-hiding security (Definition~\ref{def:indexhiding}).  

Thus we have obtained a contradiction showing that $A$ is not $(1, 1/2n)$-differentially private.  This completes the proof.
\end{proof}

\section{Cryptographic Primitives} \label{sec:cryptoprimitives}

\newcommand{\hash}{h}
\newcommand{\hashfam}{\mathcal{H}}
\newcommand{\compindist}[1]{\approx_{#1}}
\newcommand{\prffam}{\mathcal{F}}

\ifnum\lncsshort=1
We will make use of several cryptographic tools and information-theoretic primitives.  Due to space, we will omit a formal definition of standard concepts like almost-pairwise-independent hash functions, pseudorandom generators, and pseudorandom functions and defer these to the full version.
\else
\subsection{Standard Tools}
We will make use of a few standard cryptographic and information-theoretic primitives.  We will define these primitives for completeness and to set notation and terminology.
\paragraph{Almost Pairwise Independent Hash Families.}
A hash family is a family of functions $\hashfam_{s} = \set{\hash \from [s] \to \zo}$.  To avoid notational clutter, we will use the notation $\hash \getsr \hashfam$ to denote the operation of choosing a random function from $\hashfam$ and will not explicitly write the seed for the function.  We will use $|\hash|$ to denote the seed length for the function and require that $\hash$ can be evaluated in time $\poly(|\hash|)$.
\begin{definition}
A family of functions $\hashfam_{s} = \set{\hash \from [T] \to [K]}$ is \emph{$\delta$-almost pairwise independent} if for every two distinct points $x_0, x_1 \in [T]$, and every $y_0, y_1 \in [K]$,
$$
\pr{\hash \getsr \hashfam}{\hash(x_0) = y_0 \land \hash(x_1) = y_1} = \frac{1}{K^2} + \delta.
$$
\end{definition}

For every $s$, there exists a pairwise independent hash family $\hashfam_{s} = \set{\hash \from [s] \to \zo}$ such that $|\hash| = O(\log(s))$ for every $\hash \in \hashfam$.

\paragraph{Pseudorandom Generators.}
A pseudorandom generator $\prg \from \zo^{\lambda / 2} \to \zo^{\lambda}$ is a function such that
$
\prg(U_{\lambda/2}) \compindist{\negl(\lambda)} U_{\lambda}.
$
In this definition, $U_{b}$ denotes the uniform distribution on $\zo^{b}$.  Pseudorandom generators exist under the minimal assumption that one-way functions exist.

\paragraph{Pseudorandom Function Families.}
A pseudorandom function family is a family of functions $\prffam_{\lambda} = \set{\prf \from [m(\lambda)] \to [n(\lambda)]}$.  To avoid notational clutter, we will use the notation $\prf \getsr \prffam_{\lambda}$ to denote the operation of choosing a random function from $\prffam_{\lambda}$ and not explicitly write the seed for the function.  We will use $|\prf|$ to denote the description length for the function.  We require that $|\prf| = \poly(\lambda)$ and that $\prf$ can be evaluated in time $\poly(|\prf|)$.

Security requires that oracle access to $\prf \getsr \prffam_{\lambda}$ is indistinguishable from oracle access to a random function. Specifically, for all probabilistic polynomial-time algorithms $D$,
$$\left| \Pr_{\prf \getsr \prffam_{\lambda}}[D^{\prf()}(1^{\lambda}) = 1] - \Pr_{f \getsr \{f : [m] \to [n]\}}[D^{f()}(1^{\lambda}) = 1] \right| < \epsilon(\lambda)$$
for some negligible function $\epsilon$.

Under the minimal assumption that one-way functions exist, for every pair of functions $m,n$ that are at most exponential, for every $\lambda \in \N$, there is a family of pseudorandom functions $\prffam_{\lambda} = \set{\prf \from [m(\lambda)] \to [n(\lambda)]}$ such that $|\prf| = \poly(\lambda)$.

A pseudorandom function family is $\delta$-almost pairwise independent for $\delta = \negl(\lambda)$.
\fi

\subsection{Puncturable Pseudorandom Functions}
A pseudorandom function family $\prffam_{\lambda} = \set{\prf \from [m] \to [n]}$ is \emph{puncturable} if there is a deterministic procedure $\prfpuncture$ that takes as input $\prf \in \prffam_{\lambda}$ and $x^* \in [m]$ and outputs a new function $\prf^{\{x^*\}}: [m] \to [n]$ such that
\[\prf^{\{x^*\}}(x)=\begin{cases}\prf(x)&\text{if }x \neq x^* \\\bot&\text{if }x = x^*\end{cases}\]

\newcommand{\puncture}{\game{Puncture}}
The definition of security for a punctured pseudorandom function states that for any $x^*$, given the punctured function $\prf^{\{x^*\}}$, the missing value $\prf(x^*)$ is computationally unpredictable. Specifically, we define the following game {\puncture} to capture the desired security property.
\begin{figure}[h!t]
\begin{framed}
\begin{algorithmic}
\INDSTATE[0]{The challenger chooses $\prf \getsr \prffam_{\lambda}$}
\INDSTATE[0]{The challenger chooses uniform random bit $b \in \zo$, and samples $$y_0\getsr \prf(x^*),~~~~y_1 \getsr [n].$$}
\INDSTATE[0]{The challenger punctures $\prf$ at $x^*$, obtaining $\prf^{\{x^*\}}$.}
\INDSTATE[0]{The adversary is given $(y_b, \prf^{\{x^*\}})$ and outputs a bit $b'$.}
\end{algorithmic}
\end{framed}
\vspace{-6mm}
\caption{$\puncture[x^*]$}
\end{figure}

\begin{definition}[Puncturing Secure PRF]
A pseudorandom function family $\prffam_{\lambda} = \set{\prf \from [m] \to [n]}$ is \emph{$\eps$-puncturing secure} if for every $x^* \in [m]$,
$$
\pr{\puncture[x^*]}{b' = b} \leq \frac{1}{2} + \eps.
$$
\end{definition}
\newcommand{\inputmatching}{\game{InputMatching}}
\subsection{Twice Puncturable PRFs}\label{def:tpprf}

A \emph{twice puncturable PRF} is a pair of algorithms $(\prfsetup, \prfpuncture)$.
\begin{itemize}
\item $\prfsetup$ is a randomized algorithm that takes a security parameter $\lambda$ and outputs a function $\prf: [m] \to [n]$ where $m = m(\lambda)$ and $n = n(\lambda)$ are parameters of the construction.  Technically, the function is parameterized by a seed of length $\lambda$, however for notational simplicity we will ignore the seed and simply use $\prf$ to denote this function.  Formally $\prf \getsr \prfsetup(\lambda)$.

\item $\prfpuncture$ is a deterministic algorithm that takes a $\prf$ and a pair of inputs $x_0, x_1 \in [m]$ and outputs a new function $\prf^{\{x_0, x_1\}}: [m] \to [n]$ such that
\[\prf^{\{x_0,x_1\}}=\begin{cases}\prf(x)&\text{if }x\notin\{x_0,x_1\}\\\bot&\text{if }x\in\{x_0,x_1\}\end{cases}\]
Formally, $\prf^{\{x_0, x_1\}} = \prfpuncture(\prf, x_0, x_1)$.
\end{itemize}

In what follows we will always assume that $m$ and $n$ are polynomial in the security parameter and that $m = \omega(n \log(n))$.

In addition to requiring that this family of functions satisfies the standard notion of cryptographic pseudorandomness, we will now define a new security property for twice puncturable PRFs, called \emph{input matching indistinguishability}.  For any two distinct outputs $y_0,y_1\in [n],y_0 \neq y_1$, consider the following game.
\begin{figure}[h!t]
\begin{framed}
\begin{algorithmic}
\INDSTATE[0]{The challenger chooses $\prf$ such that $\forall y \in [n],\, \prf^{-1}(y) \neq \emptyset$.}
\INDSTATE[0]{The challenger chooses independent random bits $b_0, b_1 \in \zo$, and samples $$x_0\getsr \prf^{-1}(y_{b_0}),~~~~x_1 \getsr \prf^{-1}(y_{b_1}).$$}
\INDSTATE[0]{The challenger punctures $\prf$ at $x_0,x_1$, obtaining $\prf^{\{x_0,x_1\}}$.}
\INDSTATE[0]{The adversary is given $(x_0, x_1, \prf^{\{x_0, x_1\}})$ and outputs a bit $b'$.}
\end{algorithmic}
\end{framed}
\vspace{-6mm}
\caption{$\mathsf{InputMatching}[y_0, y_1]$}
\end{figure}

Notice that in this game, we have assured that every $y \in [n]$ has a preimage under $\prf$.  We need this condition to make the next step of sampling random preimages well defined.  Technically, it would suffice to have a preimage only for $y_{b_0}$ and $y_{b_1}$, but for simplicity we will assume that every possible output has a preimage.  When $f : [m] \to [n]$ is a random function, the probability that some output has no preimage is at most $n \cdot \exp(-\Omega(m/n))$ which is negligible when $m = \omega(n \log(n))$.  Since $m,n$ are assumed to be a polynomial in the security parameter, we can efficiently check if every output has a preimage, thus if $\prf$ is pseudorandom it must also be the case that every output has a preimage with high probability.  Since we can efficiently check whether or not every output has a preimage under $\prf$, and this event occurs with all but negligible probability, we can efficiently sample the pseudorandom function in the first step of ${\sf InputMatching}[y_0, y_1]$.

\begin{definition}[Input-Matching Secure PRF]
A function family $\set{\prf \from [m] \to [n]}$ is \emph{$\eps$-input-matching secure} if the function family is a secure pseudorandom function and additionally for every $y_0, y_1 \in [n]$ with $y_0 \neq y_1$,
$$
\pr{\mathsf{InputMatching}[y_0, y_1]}{b' = b_0 \oplus b_1} \leq \frac{1}{2} + \eps.
$$
\end{definition}

\ifnum\lncsshort=1
In the full version of this work we show that input-matching secure twice puncturable pseudorandom functions with suitable parameters exist.
\else
In Section~\ref{sec:prfconstruction} we will show that input-matching secure twice puncturable pseudorandom functions exist with suitable parameters.
\fi
\begin{theorem}\label{thm:existstwicepuncturableprf}
Assuming the existence of one-way functions, if $m,n$ are polynomials such that $m = \omega(n \log(n))$, then there exists a pseudorandom function family $\prffam_{\lambda} = \set{\prf\from [m(\lambda)] \to [n(\lambda)]}$ that is twice puncturable and is $\tilde{O}(\sqrt{n/m})$-input-matching secure.
\end{theorem}

\subsection{Indistinguishability Obfuscation}
We use the following formulation of Garg et al. ~\cite{GGHRSW13} for indistinguishability obfuscation:
\begin{definition}[Indistinguishability Obfuscation]
A \emph{indistinguishability obfuscator} $\obf$ for a circuit class $\{\mathcal{C}_{\lambda}\}$ is a probabilistic polynomial-time uniform algorithm satisfying the following conditions:
\begin{enumerate}
\item
$\obf(\lambda, C)$ preserves the functionality of $C$. That is, for any $C \in \mathcal{C}_{\lambda}$, if we compute $C' = \obf(\lambda, C)$, then $C'(x) = C(x)$ for all inputs $x$.

\item
For any $\lambda$ and any two circuits $C_0, C_1$ with the same functionality, the circuits $\obf(\lambda, C_0)$ and $\obf(\lambda, C_1)$ are indistinguishable. More precisely, for all pairs of probabilistic polynomial-time adversaries $(\text{Samp}, D)$, if
$$ \Pr_{(C_0, C_1, \sigma) \gets \text{Samp}(\lambda)}[(\forall x), \; C_0(x) = C_1(x)] > 1 - \negl(\lambda)$$
then 
$$| \Pr[D(\sigma, \obf(\lambda, C_0)) = 1 ]  - \Pr[D(\sigma, \obf(\lambda, C_1)) = 1 ]  | < \negl(\lambda)$$
\end{enumerate}

The circuit classes we are interested in are polynomial-size circuits - that is, when $\mathcal{C}_\lambda$ is the collection of all circuits of size at most $\lambda$. 

When clear from context, we will often drop $\lambda$ as an input to $\obf$ and as a subscript for $\mathcal{C}$.
\end{definition}

\ifnum\lncsshort=1

\section{A Weak Traitor-Tracing Scheme with Very Short Ciphertexts} \label{sec:shortctext}

In this section we construct a traitor-tracing scheme for $n$ users where the key length is polynomial in the security parameter $\lambda$ and the ciphertext length is only $O(\log(n))$.  This scheme will be used to establish our hardness result for differential privacy when the data universe can be exponentially large but the family of queries has only polynomial size.  The construction of a weak traitor-tracing scheme with user keys of length $O(\log(n))$ is in Section~\ref{sec:shortkey}.
\else
\section{A Weak Traitor-Tracing Scheme with Very Short Ciphertexts} \label{sec:shortctext}

In this section we construct a traitor-tracing scheme for $n$ users where the key length is polynomial in the security parameter $\lambda$ and the ciphertext length is only $O(\log(n))$.  This scheme will be used to establish our hardness result for differential privacy when the data universe can be exponentially large but the family of queries has only polynomial size.
\fi

\newcommand{\obfuscate}{\ensuremath{\mathsf{Obfuscate}}}

\subsection{Construction}
Let $n = \poly(\lambda)$ denote the number of users for the scheme.  Let $m = \tilde{O}(n^{\exponent})$ be a parameter.  Our construction will rely on the following primitives:
\begin{itemize}
\item A pseudorandom generator $\prg \from \zo^{\lambda/2} \to \zo^{\lambda}$.
\item A puncturable pseudorandom function family $\prffam_{\lambda, \sk} = \set{\prf_{\sk} \from [n] \to \zo^{\lambda}}$.
\item A twice-puncturable pseudorandom function family $\prffam_{\lambda, \enc} = \set{\prf_{\enc} \from [m] \to [n]}$.
\item An iO scheme $\obfuscate$.
\end{itemize}

\begin{figure}[ht!]
\begin{framed}
\begin{algorithmic}
\INDSTATE[0]{$\setup(\lambda):$}
\INDSTATE[1]{Choose $\prf_{\sk} \getsr \prffam_{\lambda, \sk}$}
\INDSTATE[1]{Choose $\prf_{\enc} \getsr \prffam_{\lambda, \enc}$ such that for every $i \in [n]$, $\prf^{-1}_{\enc}(i) \neq \emptyset$}
\INDSTATE[1]{For $i = 1,\dots,n$, let $\s_i = \prf_{\sk}(i)$.}
\INDSTATE[1]{Let $\obf \getsr \obfuscate(\pdec_{\prf_{\sk}, \prf_{\enc}})$.}
\INDSTATE[1]{Let each user's secret key be $\sk_i = (i, \s_i, \obf)$}
\INDSTATE[1]{Let the master key be $\mk = \prf_{\enc}$.}
\INDSTATE[0]{}
\INDSTATE[0]{$\enc(j, \mk = \prf_{\enc}):$}
\INDSTATE[1]{Let $c$ be chosen uniformly from $\prf_{\enc}^{-1}(j)$.}
\INDSTATE[1]{Output $c$.}
\INDSTATE[0]{}
\INDSTATE[0]{$\dec(\sk_i = (i, \s_i, \obf), c)$:}
\INDSTATE[1]{Output $\obf(c, i, \s_i)$.}
\INDSTATE[0]{}
\INDSTATE[0]{$\pdec_{\prf_{\sk}, \prf_{\enc}}(c, i, \s):$}
\INDSTATE[1]{If $\prg(\s) \neq \prg(\prf_{\sk}(i))$, halt and output $\bot$.}
\INDSTATE[1]{Output $\ind{i \leq \prf_{\enc}(c)}$.}
\end{algorithmic}
\end{framed}
\vspace{-6mm}
\caption{Our scheme $\Pi_{\mathrm{short-ctext}}$.}
\end{figure}

\begin{theorem}\label{thm:shortctext}
Assuming the existence of one-way functions and indistinguishability obfuscation.  For every polynomial $n$, the scheme $\Pi_{\mathrm{short-ctext}}$ is an $(n, d, \ell)$-traitor-tracing scheme for $d = \poly(\lambda)$ and $2^{\ell} = \tilde{O}(n^{\exponent})$ and satisfies:
\ifnum\lncsshort=1$\else$$\fi
\twoadv[i^*] \leq \frac{1}{200 n^3}.
\ifnum\lncsshort=1$\else$$\fi
\end{theorem}

Combining this theorem with Lemma~\ref{lem:2indextoindex} and Theorem~\ref{thm:tracingtodp} establishes Theorem~\ref{thm:mainshortctext} in the introduction.

\subsection*{Parameters} 

First we verify that $\Pi_\mathrm{short-ctext}$ is an $(n,d,\ell)$-traitor-tracing scheme for the desired parameters.
Observe that the length of the secret keys is $\log(n) + \lambda + |\obf|$.  By the efficiency of the pseudorandom functions and the specification of $\pdec$, the running time of $\pdec$ is $\poly(\lambda + \log(n))$.  Thus, by the efficiency of $\obfuscate$, $|\obf| = \poly(\lambda + \log(n))$.  Therefore the total key length is $\poly(\lambda + \log(n))$.  Since $n$ is assumed to be a polynomial in $\lambda$, we have that the secret keys have length $d = \poly(\lambda)$ as desired.
By construction, the ciphertext is an element of $[m]$.  Thus, since $m = \tilde{O}(n^7)$ the ciphertexts length $\ell$ satisfies $2^{\ell} = \tilde{O}(n^{\exponent})$ as desired.

\subsection{Proof of Weak Index-Hiding Security}
In light of Lemma~\ref{lem:2indextoindex}, in order to prove that the scheme satisfies weak index-hiding security, it suffices to show that for every sufficiently large $\lambda \in \N$, and every $i^* \in [n(\lambda)]$,
\ifnum\lncsshort=1$\else$$\fi
\pr{\twoindexhiding[i^*]}{b' = b_0 \oplus b_1} - \frac{1}{2}
={} o(1/n^3).
\ifnum\lncsshort=1$\else$$\fi
We will demonstrate this using a series of hybrids to reduce security of the scheme in the {\twoindexhiding} game to input-matching security of the pseudorandom function family $\prf_{\lambda, \enc}$.

Before we proceed with the argument, we remark a bit on how we will present the hybrids.  Note that the view of the adversary consists of the keys $\sk_{-i^*}$.  Each of these keys is of the form $(i, \s_i, \obf)$ where $\obf$ is an obfuscation of the same program $\pdec$.  Thus, for brevity, we will discuss only how we modify the construction of the program $\pdec$ and it will be understood that each user's key will consist of an obfuscation of this modified program.  We will also rely crucially on the fact that, because the challenge ciphertexts depend only on the master key $\mk$, we can generate the challenge ciphertexts $c_0$ and $c_1$ can be generated before the users' secret keys $\sk_1,\dots,\sk_n$.  Thus, we will be justified when we modify $\pdec$ in a manner that depends on the challenge ciphertexts and include an obfuscation of this program in the users' secret keys.  We also remark that we highlight the changes in the hybrids in {\color{DarkGreen} green}.

\subsubsection*{Breaking the decryption program for challenge index}
We use a series of hybrids to ensure that the obfuscated program reveals no information about the secret $\s_{i^*}$ for the specified user $i^*$.  First, we modify the program by hardcoding the secret $\s_{i^*}$ into the program.
\begin{figure}[h!t]
\begin{framed}
\begin{algorithmic}
\INDSTATE[0]{$\pdec^1_{\prf^{\{i^*\}}_{\sk}, \prf_{\enc}, i^*, x^*}(c, i, \s):$}
\INDSTATE[1]{\color{DarkGreen}{If $i = i^*$ and $\prg(\s) \neq x^*$}, halt and output $\bot$.}
\INDSTATE[1]{If $i \neq i^*$ and $\prg(\s) \neq \prg(\prf^{\{i^*\}}_{\sk}(i))$, halt and output $\bot$.}
\INDSTATE[1]{Output $\ind{i \leq \prf_{\enc}(c)}$.}
\end{algorithmic}
\end{framed}
\vspace{-6mm}
\caption{Modified program $\pdec^1$.  $i^*$ and $x^* = \prg(\prf_{\sk}(i^*))$ are hardcoded values.}
\end{figure}
The obfuscated versions of $\pdec$ and $\pdec^1$ are indistinguishable because the input-output behavior of the programs are identical, thus the indistinguishability obfuscation guarantees that the obfuscations of these programs are computationally indistinguishable.

\jnote{I haven't edited this stuff closely.  I am focusing more on writing up the hybrids in a notationally consistent way.}
Next we modify the setup procedure to give a uniformly random value for $s_{i^*}$.  The new setup procedure is indistinguishable from the original setup procedure by the pseudorandomness of $s_{i^*} = \prf_{\sk}(i^*)$.  Finally, we modify the decryption program to use a truly random value $x^*$ instead of $x^* = \prg(\prf_{\sk}(i^*))$.  The new decryption program is indistinguishable from the original by pseudorandomness of $\prg$ and $\prf_{\sk}$.

After making these modifications, with probability at least $1 - 2^{-\lambda/2}$, the random value $x^*$ is not in the image of $\prg$.  Thus, with probability at least $1 - 2^{-\lambda/2}$, the condition $\prg(\sk) = x^*$ will be unsatisfiable.  Therefore, we can simply remove this test without changing the program on any inputs.  Thus, the obfuscation of $\pdec^1$ will be indistinguishable from the obfuscation of the following program $\pdec^2$.

\begin{figure}[h!t]
\begin{framed}
\begin{algorithmic}
\INDSTATE[0]{$\pdec^2_{\prf^{\{i^*\}}_{\sk}, \prf_{\enc}, i^*}(c, i, \s):$}
\INDSTATE[1]{\color{DarkGreen}{If $i = i^*$ , halt and output $\bot$.}}
\INDSTATE[1]{If $i \neq i^*$ and $\prg(\s) \neq \prg(\prf^{\{i^*\}}_{\sk}(i))$, halt and output $\bot$.}
\INDSTATE[1]{Output $\ind{i \leq \prf_{\enc}(c)}$.}
\end{algorithmic}
\end{framed}
\vspace{-6mm}
\caption{Modified program $\pdec^2$.}
\end{figure}

\subsubsection*{Breaking the decryption program for the challenge ciphertexts}
First we modify the program so that the behavior on the challenge ciphertexts is hardcoded and $\prf_{\enc}$ is punctured on the challenge ciphertexts.  The new decryption program is as follows.
\begin{figure}[h!t]
\begin{framed}
\begin{algorithmic}
\INDSTATE[0]{$\pdec^3_{\prf^{\{i^*\}}_{\sk}, \prf^{\{c_0, c_1\}}_{\enc}, i^*, c_0, b_0, c_1, b_1}(c, i, \s):$}
\INDSTATE[1]{If $i = i^*$ , halt and output $\bot$.}
\INDSTATE[1]{If $i \neq i^*$ and $\prg(\s) \neq \prg(\prf^{\{i^*\}}_{\sk}(i))$, halt and output $\bot$.}
\INDSTATE[1]{{\color{DarkGreen} If $c = c_0$, output $\ind{i \leq i^* - b_0}$}}
\INDSTATE[1]{{\color{DarkGreen} If $c = c_1$, output $\ind{i \leq i^* - b_1}$}}
\INDSTATE[1]{Output $\ind{i \leq \prf^{\{c_0,c_1\}}_{\enc}(c)}$.}
\end{algorithmic}
\end{framed}
\vspace{-6mm}
\caption{Modified program $\pdec^3$.  $c_0, b_0, c_1, b_1$ are hardcoded values.}
\end{figure}
Note that the final line of the program is never reached when the input satisfies $c = c_0$ or $c = c_1$, so puncturing $\prf_{\enc}$ at these points does not affect the output of the program on any input.  Thus, $\pdec^3$ is indisintinguishable from $\pdec^2$ by the security of indistinguishability obfuscation.

Next, since, $b_0, b_1 \in \zo$, and the decryption program halts immediately if $i = i^*$, the values of $b_0, b_1$ do not affect the output of the program.  Thus, we can simply drop them from the description of the program without changing the program on any input.  So, by security of the indistinguishability obfuscation, $\pdec^3$ is indistinguishable from the following program $\pdec^4$.

\begin{figure}[h!t]
\begin{framed}
\begin{algorithmic}
\INDSTATE[0]{$\pdec^4_{\prf^{\{i^*\}}_{\sk}, \prf^{\{c_0, c_1\}}_{\enc}, i^*, c_0, c_1}(c, i, \s):$}
\INDSTATE[1]{If $i = i^*$ , halt and output $\bot$.}
\INDSTATE[1]{If $i \neq i^*$ and $\prg(\s) \neq \prg(\prf^{\{i^*\}}_{\sk}(i))$, halt and output $\bot$.}
\INDSTATE[1]{{\color{DarkGreen} If $c = c_0$, output $\ind{i \leq i^*}$}}
\INDSTATE[1]{{\color{DarkGreen} If $c = c_1$, output $\ind{i \leq i^*}$}}
\INDSTATE[1]{Output $\ind{i \leq \prf^{\{c_0,c_1\}}_{\enc}(c)}$.}
\end{algorithmic}
\end{framed}
\vspace{-6mm}
\caption{Modified program $\pdec^4$.  $c_0, c_1$ are hardcoded values.}
\end{figure}

\subsubsection*{Reducing to Input-Matching Security}
Finally, we claim that if the adversary is able to win at {\twoindexhiding} then he can also win the game $\inputmatching[i^*-1, i^*]$, which violates input-matching security of $\prffam_{\lambda, \enc}$.  

Recall that the challenge in the game $\inputmatching[i^*-1,i^*]$ consists of a tuple $(c_0, c_1, \prf^{\{c_0, c_1\}})$ where $\prf_{\enc}$ is sampled subject to 1) $\prf_{\enc}(c_0) = i^* - b_0$ for a random $b_0 \in \zo$, 2) $\prf_{\enc}(c_1) = i^* - b_1$ for a random $b_1 \in \zo$, and 3) $\prf_{\enc}^{-1}(i) \neq \emptyset$ for every $i \in [n]$.  Given this input, we can precisely simulate the view of the adversary in $\twoindexhiding[i^*]$.  To do so, we can choose $\prf_{\sk}$ and give the keys $\sk_{-i^*}$ and obfuscations of $\pdec^4_{\prf^{\{i^*\}}_{\sk}, \prf^{\{c_0, c_1\}}_{\enc}, i^*, c_0, c_1}$ to the adversary.  Then we can user $c_0, c_1$ as the challenge ciphertexts and obtain a bit $b'$ from the adversary.  By input-matching security, we have that
\ifnum\lncsshort=1$\else$$\fi
\pr{}{b' = b_0 \oplus b_1} - \frac{1}{2}
={} o(1/n^3).
\ifnum\lncsshort=1$\else$$\fi
Since, as we argued above, the view of the adversary in this game is indistinguishable from the view of the adversary in $\twoindexhiding[i^*]$, we conclude that
\ifnum\lncsshort=1$\else$$\fi
\pr{\twoindexhiding[i^*]}{b' = b_0 \oplus b_1} - \frac{1}{2}
={} o(1/n^3),
\ifnum\lncsshort=1$\else$$\fi
as desired.  This completes the proof.

\section{A Weak Traitor-Tracing Scheme with Very Short Keys} \label{sec:shortkey}
\ifnum\lncsshort=1
In the full version of this work, we construct a different traitor-tracing scheme for $n$ users where the length of the secret user keys is $O(\log(n))$ and the length of the ciphertexts is $\poly(\lambda)$, which is used to establish Theorem~\ref{thm:mainshortkey}.

The construction is essentially ``dual'' to short-ciphertext construction from Section~\ref{sec:shortctext}.  At a very high level: in the short-ciphertext construction, the ciphertext represented a (short) encrypted index and the user keys consisted of a (short) user id, a (long) secret key, and a (long) obfuscated program that takes the encrypted index, decrypts it, and decides if that user is meant to output $0$ or $1$.  In the short-key construction, the user key consists of a (short) user id and a (short) secret key, and the ciphertext is a (long) obfuscated program that contains an index, takes a user id and a key, and decides if that user is meant to output $0$ or $1$.  The main challenge in the analysis is that, because the secret keys are short, the adversary can query the obfuscated program on all (user id, secret key) pairs, including those that it doesn't know.  Thus, we need to design the behavior of the obfuscated program in such a way that the adversary can't distinguish valid and invalid (user id, secret key) pairs.  That is, it can see the behavior of the obfuscated program on all inputs, but it doesn't know \emph{which} input corresponds to the information it wants to learn.

\else
In this section we construct a different traitor-tracing scheme for $n$ users where the parameters are essentially reversed---the length of the secret user keys is $O(\log(n))$ and the length of the ciphertexts is $\poly(\lambda)$.  This scheme will be used to establish our hardness result for differential privacy when the number of queries is exponentially large but the data universe has only polynomial size.

\subsection{Construction}
Let $n = \poly(\lambda)$ denote the number of users for the scheme.  Let $m = \tilde{O}(n^{\otherexponent})$ be a parameter.  Our construction will rely on the following primitives:
\begin{itemize}
\item A puncturable pseudorandom function family $\prffam_{\lambda, \sk} = \set{\prf_\sk \from [n] \to [m]}$.
\item A puncturable pseudorandom function family $\prffam_{\lambda, \enc} = \set{\prf_\enc \from [n] \times [m] \to \zo}$.
\item An iO scheme $\obfuscate$.
\end{itemize}

\begin{figure}[h!t]
\begin{framed}
\begin{algorithmic}
\INDSTATE[0]{$\setup(\lambda):$}
\INDSTATE[1]{Choose a pseudorandom function $\prf_{\sk} \getsr \prffam_{\lambda, \sk}$.}
\INDSTATE[1]{For $i = 1,\dots,n$, let $\s_i = \prf_{\sk}(i)$, and let each user's secret key be $\sk_i = (i, \s_i) \in [n] \times [m]$.}
\INDSTATE[1]{Let the master key be $\mk = \prf_{\sk}$.}
\INDSTATE[0]{}
\INDSTATE[0]{$\enc(j, \mk = \prf_{\sk}):$}
\INDSTATE[1]{Choose a pseudorandom function $\prf_\enc \getsr \prffam_{\lambda,\enc}$.}
\INDSTATE[1]{Let $\obf = \obfuscate(\pdec_{j, \prf_{\sk}, \prf_\enc})$}
\INDSTATE[1]{Output $c = \obf$.}
\INDSTATE[0]{}
\INDSTATE[0]{$\dec(\sk_i = (i, \s_i), c = \obf)$:}
\INDSTATE[1]{Output $\obf(i, \sk_i)$.}
\INDSTATE[0]{}
\INDSTATE[0]{$\pdec_{j, \prf_{\sk}, \prf_\enc}(i, \s)$:}
\INDSTATE[1]{If $\s \neq \prf_{\sk}(i)$, output $\prf_\enc(i,\s)$.}
\INDSTATE[1]{Else, output $\ind{i \leq j}$.}
\end{algorithmic}
\end{framed}
\vspace{-6mm}
\caption{Our scheme $\Pi_{\mathrm{short-key}}$}
\end{figure}

\begin{theorem}\label{thm:shortkey}
Assuming the existence of one-way functions and indistinguishability obfuscation, for every polynomial $n$, the scheme $\Pi_{\mathrm{short-key}}$ is an $(n, d, \ell)$-traitor-tracing scheme for $2^{d} = \tilde{O}(n^{\exponent})$ and $\ell = \poly(\lambda)$, and is weakly index-hiding secure.
\end{theorem}

Combining this theorem with Lemma~\ref{lem:2indextoindex} and Theorem~\ref{thm:tracingtodp} establishes Theorem~\ref{thm:mainshortkey} in the introduction.

\subsection*{Parameters} 

First we verify that $\Pi_\mathrm{short-key}$ is an $(n,d,\ell)$-traitor-tracing scheme for the desired parameters.
Observe that the length of the secret keys is $d$ such that $2^d = n m$.  By construction, since $m = \tilde{O}(n^{6})$, $2^d = \tilde{O}(n^{\exponent})$.  The length of the ciphertext is $|\obf|$, which is $\poly(|\pdec|)$ by the efficiency of the obfuscation scheme.  By the efficiency of the pseudorandom function family and the pairwise independent hash family, the running time of $\pdec$ is at most $\poly(\lambda + \log(n))$.  Since $n$ is assumed to be a polynomial in $\lambda$, the ciphertexts have length $\poly(\lambda)$.

\subsection{Proof of Weak Index-Hiding Security}
Just as in Section~\ref{sec:shortctext}, we will rely on Lemma~\ref{lem:2indextoindex} so that we only need to show that for every $\lambda \in \N$, and every $i^* \in [n(\lambda)]$,
$$
\pr{\twoindexhiding[i^*]}{b' = b_0 \oplus b_1} - \frac{1}{2}
={} o(1/n^3).
$$
We will demonstrate this using a series of hybrids to reduce security of the scheme in the {\twoindexhiding} game to the security of the pseudorandom function families.

In our argument, recall that the adversary's view consists of the keys $\sk_{-i^*}$ and the challenge ciphertexts $c_0, c_1$.  In our proof, we will not modify how the keys are generated, so we will present the hybrids only by how the challenge ciphertexts are generated.  Also, for simplicity, we will focus only on how $c_0$ is generated as a function of $i^*, b_0$ and $\mk$.  The ciphertext $c_1$ will be generated in exactly the same way but as a function of $i^*, b_1$ and $\mk$.  We also remark that we highlight the changes in the hybrids in {\color{DarkGreen} green}.

\subsubsection*{Hiding the missing user key}
First we modify the encryption procedure to one where $\prf_{\sk}$ is punctured on $i^*$ and the value $\s^* = \prf_{\sk}(i^*)$ is hardcoded into the program.
\begin{figure}[h!t]
\begin{framed}
\begin{algorithmic}
\INDSTATE[0]{$\enc^1(i^*, b_0, \mk = \prf_{\sk}):$}
\INDSTATE[1]{Choose a pseudorandom function $\prf_\enc \getsr \prffam_{\lambda,\enc}$.}
\INDSTATE[1]{{\color{DarkGreen} Let $\s^* = \prf_{\sk}(i^*)$, $\prf_{\sk}^{\{i^*\}} = \prfpuncture(\prf_{\sk}, i^*)$.}}
\INDSTATE[1]{$$\textrm{Let}~\obf = \obfuscate\left({\color{DarkGreen}\pdec^{1}_{i^*, b_0, \s^*, \prf^{\{i^*\}}_{\sk}, \prf_\enc}}\right).$$}
\INDSTATE[1]{Output $c_0 = \obf$.}
\INDSTATE[0]{}
\INDSTATE[0]{$\pdec^{1}_{i^*, b_0, \s^*, \prf_{\sk}^{\{i^*\}}, \prf_\enc}(i, \s)$:}
\INDSTATE[1]{{\color{DarkGreen} If $i = i^*$}}
\INDSTATE[2]{{\color{DarkGreen} If $\s \neq \s^*$, output $\prf_\enc(i^*,\s)$}}
\INDSTATE[2]{{\color{DarkGreen} If $\s = \s^*$, output $1-b_0$}}
\INDSTATE[1]{Else If $i \neq i^*$}
\INDSTATE[2]{If $\s \neq \prf^{\{i^*\}}_{\sk}(i)$, halt and output $\prf_\enc(i,\s)$.}
\INDSTATE[2]{\color{DarkGreen} Output $\ind{i \leq i^* - 1}$.}
\end{algorithmic}
\end{framed}
\vspace{-6mm}
\caption{Hybrid $(\enc^1, \pdec^1)$.}
\end{figure}

We claim that, by the security of the iO scheme, the distribution of $c_0, c_1$ under $\enc^1$ is computationally indistinguishable from the distribution of $c_0, c_1$ under $\enc$.  The reason is that the obfuscation $\pdec$ and $\pdec^1$ compute the same function.  Consider two cases, depending on whether $i = i^*$ or $i \neq i^*$.  If $i \neq i^*$, since $b_0 \in \zo$, and $i \neq i^*$, replacing $\mathbb{I}\{i \leq i^* - b_0\}$ with $\mathbb{I}\{i \leq i^* - 1\}$ does not change the output.  Moreover, since we only reach the branch involving $\prf_{\sk}^{\{i^*\}}$ when $i \neq i^*$, the puncturing does not affect the output of the program.  If $i = i^*$, then the program either outputs $\prf_\enc(i^*, s)$ as it did before when $s \neq s^*$ or it outputs $1 - b_0$: equivalent to $\ind{i \leq i^* - b_0}$. Thus, by iO, the obfuscated programs are indistinguishable.


Next, we argue that, since $\prf_{\sk}^{\{i^*\}}$ is sampled from a puncturable pseudorandom function family, and the adversary's view consists of $\s_{-i^*} = \{\prf_{\sk}(i)\}_{i \neq i^*}$ but not $\prf_{\sk}(i^*)$, the value of $\prf_{\sk}(i^*)$ is computationally indistinguishable to the adversary from a random value.  Thus, we can move to another hybrid $(\enc^2, \pdec^2)$ where the value $\s^*$ is replaced with a uniformly random value $\tilde{s}$.

\begin{figure}[h!t]
\begin{framed}
\begin{algorithmic}
\INDSTATE[0]{$\enc^2(i^*, b_0, \mk = \prf_{\sk}):$}
\INDSTATE[1]{Choose a pseudorandom function $\prf_\enc \getsr \prffam_{\lambda,\enc}$.}
\INDSTATE[1]{$\prf_{\sk}^{\{i^*\}} = \prfpuncture(\prf_{\sk}, i^*)$, {\color{DarkGreen} Let $\tilde{s} \getsr [m]$.}}
\INDSTATE[1]{$$\textrm{Let}~\obf = \obfuscate\left({\color{DarkGreen}\pdec^{2}_{i^*, b_0, \tilde{s}, \prf^{\{i^*\}}_{\sk}, \prf_\enc}}\right).$$}
\INDSTATE[1]{Output $c_0 = \obf$.}
\INDSTATE[0]{}
\INDSTATE[0]{$\pdec^{2}_{i^*, b_0, \tilde{s}, \prf_{\sk}^{\{i^*\}}, \prf_\enc}(i, \s)$:}
\INDSTATE[1]{If $i = i^*$}
\INDSTATE[2]{{\color{DarkGreen} If $\s \neq \tilde{s}$, output $\prf_\enc(i^*,\s)$}}
\INDSTATE[2]{\color{DarkGreen}  If $\s = \tilde{s}$, output $1-b_0$}
\INDSTATE[1]{Else If $i \neq i^*$}
\INDSTATE[2]{If $\s \neq \prf^{\{i^*\}}_{\sk}(i)$, halt and output $\prf_\enc(i,\s)$.}
\INDSTATE[2]{Output $\ind{i \leq i^* - 1}$.}
\end{algorithmic}
\end{framed}
\vspace{-6mm}
\caption{Hybrid $(\enc^2, \pdec^2)$.}
\end{figure}

\paragraph{Hiding the challenge index}
Now we want to remove any explicit use of $b_0$ from $\pdec^2$.  The natural way to try to do this is to remove the line where the program outputs $1 - b_0$ when the input is $(i^*, \tilde{s})$, and instead have the program output $\prf_\enc(i^*, \tilde{s})$.  However, this would involve changing the program's output on one input, and indistinguishability obfuscation does not guarantee any security in this case.  We get around this problem in two steps.  First, we note that the value of $\prf_\enc$ on the point $(i^*,\tilde{s})$ is never needed in $\pdec^2$, so we can move to a new procedure $\pdec^3$ where we puncture at that point without changing the program functionality.  Indistinguishability obfuscation guarantees that $\pdec^2$ and $\pdec^3$ are computationally indistinguishable.

\begin{figure}[h!t]
	\begin{framed}
		\begin{algorithmic}
			\INDSTATE[0]{$\enc^3(i^*, b_0, \mk = \prf_{\sk}):$}
			\INDSTATE[1]{\color{DarkGreen} Let $\tilde{s} \getsr [m]$.}
			\INDSTATE[1]{Choose a pseudorandom function $\prf_\enc \getsr \prffam_{\lambda,\enc}$}
			\INDSTATE[1]{$\color{DarkGreen} \prf_\enc^{\{(i^*,\tilde{s})\}}=\prfpuncture{\prf_\enc,(i^*,\tilde{s})}$.}
			\INDSTATE[1]{$\prf_{\sk}^{\{i^*\}} = \prfpuncture(\prf_{\sk}, i^*)$.}
			\INDSTATE[1]{$$\textrm{Let}~\obf = \obfuscate\left({\color{DarkGreen}\pdec^{3}_{i^*, b_0, \tilde{s}, \prf^{\{i^*\}}_{\sk}, \prf_\enc^{\{(i^*,\tilde{s})\}}}}\right).$$}
			\INDSTATE[1]{Output $c_0 = \obf$.}
			\INDSTATE[0]{}
			\INDSTATE[0]{$\pdec^{3}_{i^*, b_0, \tilde{s}, \prf_{\sk}^{\{i^*\}}, \prf_\enc^{\{(i^*,\tilde{s})\}}}(i, \s)$:}
			\INDSTATE[1]{If $i = i^*$}
			\INDSTATE[2]{{\color{DarkGreen} If $\s \neq \tilde{s}$, output $\prf_\enc^{\{(i^*,\tilde{s})\}}(i^*,\s)$}}
			\INDSTATE[2]{If $\s = \tilde{s}$, output $1-b_0$}
			\INDSTATE[1]{Else If $i \neq i^*$}
			\INDSTATE[2]{If $\s \neq \prf^{\{i^*\}}_{\sk}(i)$, halt and output $\prf_\enc^{\{(i^*,\tilde{s})\}}(i,\s)$.}
			\INDSTATE[2]{Output $\ind{i \leq i^* - 1}$.}
		\end{algorithmic}
	\end{framed}
	\vspace{-6mm}
	\caption{Hybrid $(\enc^3, \pdec^3)$.}
\end{figure}

Next, we define another hybrid $\pdec^4$ where change how we sample $\prf_\enc$ and sample it so that $\prf_\enc(i^*,\tilde{s})=1-b_0$.  Observe that the hybrid only depends on $\prf_\enc^{\{(i^*,\tilde{s})\}}$.  We claim the distributions of $\prf_\enc^{\{(i^*,\tilde{s})\}}$ when $\prf_\enc$ is sampled correctly versus sampled conditioned on $\prf_\enc(i^*,\tilde{s})=1-b_0$ are computationally indistinguishable. This follows readily from punctured PRF security.  Suppose to the contrary that the two distributions were distinguishable with non-negligible advantage $\delta$ by adversary $A$.  Then consider a punctured PRF adversary $B$ that is given $\prf_\enc^{\{(i^*,\tilde{s})\}},b$ where $b$ is chosen at random, or $b=\prf_\enc(i^*,\tilde{s})$.  $B$ distinguishes the two cases as follows.  If $b\neq 1-b_0$, then $B$ outputs a random bit and stops.  Otherwise, it runs $A$ on $\prf_\enc^{\{(i^*,\tilde{s})\}}$, and outputs whatever $A$ outputs.  If $b$ is truly random and independent of $\prf_\enc$, then conditioned on $b=1-b_0$, $\prf_\enc$ is sampled randomly.  However, if $b=\prf_\enc(i^*,\tilde{s})$, then conditioned on $b=1-b_0$, $\prf_\enc$ is sampled such that $\prf_\enc(i^*,\tilde{s})=1-b_0$.  These are exactly the two cases that $A$ distinguishes.  Hence, conditioned on $b=1-b_0$, $B$ guesses correctly with probability $\frac{1}{2}+\delta$.  Moreover, by PRF security, $b=1-b_0$ with probability $\geq \frac{1}{2} - \eps$ for some negligible quantity $\eps$, and in the case $b\neq 1-b_0$, $B$ guess correctly with probability $\frac{1}{2}$.  Hence, overall $B$ guesses correctly with probability $\geq \frac{1}{2} (\frac{1}{2} + \eps) +(\frac{1}{2}+\delta)(\frac{1}{2} - \eps) = \frac{1}{2} + \frac{\delta}{2} - \eps \delta$.  Hence, $B$ has non-negligible advantage $ \frac{\delta}{2} - \eps \delta$.  Thus, changing how $\prf_\enc$ is sampled is computationally undetectable, and $\pdec$ is otherwise unchanged.  Therefore $\pdec^3$ and $\pdec^4$ are computationally indistinguishable.

\begin{figure}[h!t]
	\begin{framed}
		\begin{algorithmic}
			\INDSTATE[0]{$\enc^4(i^*, b_0, \mk = \prf_{\sk}):$}
			\INDSTATE[1]{Let $\tilde{s} \getsr [m]$.}
			\INDSTATE[1]{Choose a pseudorandom function $\prf_\enc \getsr \prffam_{\lambda,\enc}$ {\color{DarkGreen} conditioned on $\prf_\enc(i^*,\tilde{s})=1-b_0$}.}
			\INDSTATE[1]{$\prf_\enc^{\{(i^*,\tilde{s})\}}=\prfpuncture{\prf_\enc,(i^*,\tilde{s})}$.}
			\INDSTATE[1]{$\prf_{\sk}^{\{i^*\}} = \prfpuncture(\prf_{\sk}, i^*)$.}
			\INDSTATE[1]{$$\textrm{Let}~\obf = \obfuscate\left(\pdec^{4}_{i^*, b_0, \tilde{s}, \prf^{\{i^*\}}_{\sk}, \prf_\enc^{\{(i^*,\tilde{s})\}}}\right).$$}
			\INDSTATE[1]{Output $c_0 = \obf$.}
			\INDSTATE[0]{}
			\INDSTATE[0]{$\pdec^{4}_{i^*, b_0, \tilde{s}, \prf_{\sk}^{\{i^*\}}, \prf_\enc^{\{(i^*,\tilde{s})\}}}(i, \s)$:}
			\INDSTATE[1]{If $i = i^*$}
			\INDSTATE[2]{If $\s \neq \tilde{s}$, output $\prf_\enc^{\{(i^*,\tilde{s})\}}(i^*,\s)$}
			\INDSTATE[2]{If $\s = \tilde{s}$, output $1-b_0$}
			\INDSTATE[1]{Else If $i \neq i^*$}
			\INDSTATE[2]{If $\s \neq \prf^{\{i^*\}}_{\sk}(i)$, halt and output $\prf_\enc^{\{(i^*,\tilde{s})\}}(i,\s)$.}
			\INDSTATE[2]{Output $\ind{i \leq i^* - 1}$.}
		\end{algorithmic}
	\end{framed}
	\vspace{-6mm}
	\caption{Hybrid $(\enc^4, \pdec^4)$.}
\end{figure}

Next, since $\prf_\enc(i^*,\tilde{s})=1-b_0$, we can move to another hybrid $\pdec^5$ where we delete the line ``If $\s = \tilde{s}$, output $1-b_0$'' without changing the functionality.  Thus, by indistinguishability obfuscation, $\pdec^4$ and $\pdec^5$ are computationally indistinguishable.

\begin{figure}[h!t]
	\begin{framed}
		\begin{algorithmic}
			\INDSTATE[0]{$\enc^5(i^*, b_0, \mk = \prf_{\sk}):$}
			\INDSTATE[1]{Let $\tilde{s} \getsr [m]$.}
			\INDSTATE[1]{Choose a pseudorandom function $\prf_\enc \getsr \prffam_{\lambda,\enc}$ such that $\prf_\enc(i^*,\tilde{s})=1-b_0$}
			\INDSTATE[1]{$\prf_{\sk}^{\{i^*\}} = \prfpuncture(\prf_{\sk}, i^*)$.}
			\INDSTATE[1]{$$\textrm{Let}~\obf = \obfuscate\left({\color{DarkGreen}\pdec^{5}_{i^*, \prf^{\{i^*\}}_{\sk}, \prf_\enc}}\right).$$}
			\INDSTATE[1]{Output $c_0 = \obf$.}
			\INDSTATE[0]{}
			\INDSTATE[0]{$\pdec^{5}_{i^*, \prf_{\sk}^{\{i^*\}}, \prf_\enc}(i, \s)$:}
			\INDSTATE[1]{If $i = i^*$}
			\INDSTATE[2]{Output $\prf_\enc(i^*,\s)$}
			\INDSTATE[1]{Else If $i \neq i^*$}
			\INDSTATE[2]{If $\s \neq \prf^{\{i^*\}}_{\sk}(i)$, halt and output $\prf_\enc(i,\s)$.}
			\INDSTATE[2]{Output $\ind{i \leq i^* - 1}$.}
		\end{algorithmic}
	\end{framed}
	\vspace{-6mm}
	\caption{Hybrid $(\enc^5, \pdec^5)$.}
\end{figure}

Now notice that $\pdec^5$ is independent of $b_0$.  However, $\enc^5$ still depends on $b_0$.  We now move to the final hybrid $\pdec^6$ where we remove the condition that $\prf_\enc(i^*,\tilde{s})=1-b_0$, which will completely remove the dependence on $b_0$.  

\begin{figure}[h!t]
	\begin{framed}
		\begin{algorithmic}
			\INDSTATE[0]{$\enc^6(i^*, \mk = \prf_{\sk}):$}
			\INDSTATE[1]{Choose a pseudorandom function $\prf_\enc \getsr \prffam_{\lambda,\enc}$}
			\INDSTATE[1]{$\prf_{\sk}^{\{i^*\}} = \prfpuncture(\prf_{\sk}, i^*)$.}
			\INDSTATE[1]{$$\textrm{Let}~\obf = \obfuscate\left(\pdec^{6}_{i^*, \prf^{\{i^*\}}_{\sk}, \prf_\enc}\right).$$}
			\INDSTATE[1]{Output $c_0 = \obf$.}
			\INDSTATE[0]{}
			\INDSTATE[0]{$\pdec^{6}_{i^*, \prf_{\sk}^{\{i^*\}}, \prf_\enc}(i, \s)$:}
			\INDSTATE[1]{If $i = i^*$}
			\INDSTATE[2]{Output $\prf_\enc(i^*,\s)$}
			\INDSTATE[1]{Else If $i \neq i^*$}
			\INDSTATE[2]{If $\s \neq \prf^{\{i^*\}}_{\sk}(i)$, halt and output $\prf_\enc(i,\s)$.}
			\INDSTATE[2]{Output $\ind{i \leq i^* - 1}$.}
		\end{algorithmic}
	\end{framed}
	\vspace{-6mm}
	\caption{Hybrid $(\enc^6, \pdec^6)$.}
\end{figure}

To prove that $\enc^6$ is indistinguishable from $\enc^5$, notice that they are independent of $\tilde{s}$, except through the sampling of $\prf_\enc$.  Using this, and the following lemma, we argue that we can remove the condition that $\prf_\enc(i^*,\tilde{s})=1-b_0$.  

\begin{lemma} \label{lem:conditionalhashlemma}
Let $\hashfam = \set{\hash \from [T] \to [K]}$ be a $\delta$-almost pairwise independent hash family.  Let $y \in [K]$ and $M \subseteq [T]$ of size $m$ be arbitrary.  Define the following two distributions.  
\begin{itemize}
	\item $D_1$: Choose $\hash \getsr \hashfam$.
	\item $D_2$: Choose a random $x \in M$, and then choose $\hash \getsr (\hashfam \mid \hash(x) = y)$.
\end{itemize}
Then $D_1$ and $D_2$ are $(\frac{1}{2}\sqrt{K/m+7K^2 \delta})$-close in statistical distance.
\end{lemma}

\ifnum\lncsshort=0
We defer the proof to Section~\ref{sec:proofofconditionalhashlemma}.  
\else
We defer the proof to the full version.
\fi
The natural way to try to show that $(\enc^6, \pdec^6)$ is $o(1/n^3)$ statistically close to $(\enc^5, \pdec^5)$ is to apply this lemma to the hash family $\hashfam=\prffam_{\lambda,\enc}$.  Recall that a pseudorandom function family is also $\negl(\lambda)$-pairwise independent.  Here, the parameters would be $[T] = [n] \times [m]$, $M = \set{(i^*, s) \mid s \in [m]}$ and $b = 1-b_0$, and the random choice $x \in M$ is the pair $(i^*, \tilde{s})$.  

However, recall that the adversary not only sees $c_0 = \enc^5(i^*,b_0,\mk)$, but also sees $c_1 = \enc^5(i^*,b_1,\mk)$, and these share the same $\tilde{s}$.  Hence, we cannot directly invoke Lemma~\ref{lem:conditionalhashlemma} on the $\prf_{\enc,0}$ sampled in $c_0$, since $\tilde{s}$ is also used to sample $\prf_{\enc,1}$ when sampling $c_1$, and is therefore not guaranteed to be random given $c_1$.  

Instead, we actually consider the function family $\hashfam=\prffam_{\lambda,\enc}^2$, where we define $$h(i,s)=(\prf_{\enc,0},\prf_{\enc,1})(i,s)=(\prf_{\enc,0}(i,s),\prf_{\enc, 1}(i,s)).$$  In $\enc^5$, $h$ is drawn at random conditioned on $h(i^*,\tilde{s})=(1-b_0,1-b_1)$, whereas in $\enc^6$, it is drawn at random.

$\hashfam$ is still a pseudorandom function family, so it must be $\negl(\lambda)$-almost pairwise independent with $\delta$ negligible.  In particular, $\delta=o(1/m)$.  Hence, the conditions of Lemma~\ref{lem:conditionalhashlemma} are satisfied with $K=4$.  Since the description of $\pdec^5, \pdec^6$ is the tuple $(i^*, \tilde{s}, \prf_{\sk}^{\{i^*\}}, \prf_{\enc,0},\prf_{\enc,1})$, and by Lemma~\ref{lem:conditionalhashlemma} the distribution on these tuples differs by at most $O(\sqrt{1/m})$ in statistical distance, we also have that the distribution on obfuscations of $\pdec^5, \pdec^6$ differs by at most $O(\sqrt{1/m})$.  Finally, we can choose a value of $m = \tilde{O}(n^{6})$ so that $O(\sqrt{1/m}) = o(1/n^3)$.

Observe that when we generate user keys $\sk_{-i^*}$ and the challenge ciphertexts according to $(\enc^6, \pdec^6)$, the distribution of the adversary's view is completely independent of the random values $b_0, b_1$.  Thus no adversary can output $b' = b_0 \oplus b_1$ with probability greater than $1/2$.  Since the distribution of these challenge ciphertexts is $o(1/n^3)$-computationally indistinguishable from the original distribution on challenge ciphertexts, we have that for every computationally efficient adversary,
$$
\pr{\twoindexhiding[i^*]}{b' = b_0 \oplus b_1} - \frac{1}{2} = o(1/n^3),
$$
as desired.  This completes the proof.

\ifnum\lncsshort=0
\subsection{Proof of Lemma~\ref{lem:conditionalhashlemma}} \label{sec:proofofconditionalhashlemma}
We will fix $y = 1$ for simplicity.  The cases of $y = 2,\dots,K$ follow symmetrically.

We will first bound the R\'{e}nyi divergence between $D_1$ and $D_2$, which is defined as
$$
\mathit{RD}(D_1,D_2)=\sum_\hash \frac{\prob{H = \hash:H \getsr D_2}^2}{\prob{H = \hash:H\getsr D_1}}
$$
Here, $\hash$ ranges over the support of $D_2$ (since the support of $D_2$ is a subset of $H$, we can equivalently view the sum as one over all $h$ in $H$).  Once we do this, we will obtain an upper bound on the statistical distance between $D_1$ and $D_2$ using the inequality
\begin{equation}\label{eq:RDtoSD}
\mathit{SD}(D_1,D_2)\leq\frac{\sqrt{\mathit{RD}(D_1,D_2)-1}}{2}.
\end{equation}

To bound the R\'{e}nyi divergence, we can start by writing
\begin{align*}
\prob{H = \hash:H \gets D_2}^2&=\left(\frac{1}{m}\sum_{x \in M}\prob{H = \hash:H(x)=1}\right)^2\\
&= \frac{1}{m^2}\sum_{x,x' \in M}\prob{H = \hash:H(x)=1}\pr{}{H = \hash:H(x')=1}
\end{align*}
Where in all the (conditional) probabilities on the right, $\hash$ is drawn from $D_1$, conditioned on some event.  
This allows us to write
$$
\mathit{RD}(D_1,D_2)=\frac{1}{m^2}\sum_{x,x' \in M}\sum_\hash \frac{\prob{H = \hash : H(x)=1}\prob{H = \hash:H(x')=1}}{\prob{H = \hash}}
$$

We now divide the sum into two cases.
\begin{itemize}
	\item $x=x'$.  In this case, the summand becomes $\prob{H=\hash:H(x)=1}^2/\prob{H=\hash}$.  Notice that
	\begin{equation*}
	\prob{H=\hash:H(x)=1}=
	\begin{cases}
	0&\text{if }h(x)\neq 1\\
	\frac{\prob{H=\hash}}{\prob{H(x)=1}}&\text{if }h(x) = 1
	\end{cases}
	\end{equation*}
		
	Therefore, the summand is 
	\begin{equation*}
	\frac{\prob{H=\hash:H(x)=1}^2}{\prob{H=\hash}}=\begin{cases}0&\text{if }h(x)\neq 1\\\frac{\prob{H=\hash}}{\prob{H(x)=1}^2}&\text{if }h(x)=1\end{cases}
	\end{equation*}
	
	Now, notice that $\sum_{\hash:\hash(x)=1}\prob{H=\hash}=\prob{H(x)=1}$.  Thus, if we carry out the sum over $\hash$, the summand becomes $1/\prob{H(x)=1}$.  
	
	\item $x\neq x'$.  Then 
	\begin{align*}
	\prob{H=\hash:H(x)=1}={}&\prob{H=\hash:H(x)=1,H(x')\neq1}\prob{H(x')\neq1:H(x)=1}+\\&\prob{H=\hash:H(x)=1,H(x')=1}\prob{H(x')=1:H(x)=1}\\
	\prob{H=\hash:H(x')=1}={}&\prob{H=\hash:H(x')=1, H(x)\neq1}\prob{H(x)\neq1:H(x')=1}+\\&\prob{H=\hash:H(x')=1,H(x)=1}\prob{H(x)=1:H(x')=1}
	\end{align*}
	
	When we take the product of the two expressions and expand, we obtain four products, only one of which is nonzero (when $\hash(x)=\hash(x')=1$):
	\begin{align*}
	&\prob{H=\hash:H(x)= 1,H(x') = 1}^{2} \cdot \prob{H(x')= 1:H(x)= 1} \cdot \prob{H(x)= 1:H(x')= 1}
	\end{align*}
	
	Therefore, the summand is 
	\begin{align*}
	&\frac{\prob{H=\hash:H(x)=1}\prob{H=\hash:H(x')=1}}{\prob{H=\hash}} \\
	={} &\frac{\prob{H(x')=1:H(x)=1}\prob{H(x)=1:H(x')=1}}{\prob{H=\hash}}\cdot \prob{H=\hash:H(x)=1,H(x')=1}^2 \\
	={} &\frac{\prob{H = \hash}}{\prob{H(x)=1}\prob{H(x')=1}} \cdot \ind{\hash(x) = \hash(x') = 1}
	\end{align*}
	
	When we sum over all $\hash$, we get 
	$$
	\sum_{\hash} \frac{\prob{H=\hash:H(x)=1}\prob{H=\hash:H(x')=1}}{\prob{H=\hash}}=\frac{\prob{H(x)=1\wedge H(x')=1}}{\prob{H(x)=1}\prob{H(x')=1}}
	$$
\end{itemize}

Therefore, the R\'{e}nyi divergence is
$$
\mathit{RD}(D_1,D_2)=\frac{1}{m^2}\left(\left(\sum_{x \in M} \frac{1}{\prob{H(x)=1}}\right)+\left(\sum_{x\neq x' \in M}\frac{\prob{H(x)=1\wedge H(x')=1}}{\prob{H(x)=1}\prob{H(x')=1}}\right)\right)$$

We now invoke $\delta$-almost pairwise independence to claim that
\begin{itemize}
	\item For every $x\in M$, $\prob{H(x) = 1} \geq 1/K-\delta$. 
	\item For every $x\neq x' \in M$, $\prob{H(x) = 1 \land H(x') = 1} \leq 1/K^2+\delta$.
\end{itemize}

Therefore, for $\delta\leq 1/2K$, it is easy to show that $1/\prob{H(x)=1} \leq 2+2K^2\delta$ and $$\frac{\prob{H(x)=1\wedge H(x')=1}}{\prob{H(x)=1}\prob{H(x')=1}} \leq 1+7K^2\delta.$$  So we have that
$$
\mathit{RD}(D_1,D_2)\leq \frac{1}{m^2}\left((K+2K^2\delta)m+(m^2 - m)(1+7K^2 \delta)\right) \leq 1 + \frac{K-1}{m} + 7K^2\delta.
$$
Using the relationship between statistical distance and R\'{e}nyi divergence above (Equation~\eqref{eq:RDtoSD}), we obtain $\mathit{SD}(D_1,D_2) \leq \frac{1}{2}\sqrt{(K-1)/m+7K^2\delta}$, as long as $\delta<1/2k$.  Notice that for $\delta\geq 1/2K$, $7K^2\delta\geq 7$, and so our bound is larger than 1 anyway.  Hence, the bound holds for all $\delta$.
\fi
\fi

\ifnum\lncs=0
\subsection*{Acknowledgments}
The first author is supported by an NSF Graduate Research Fellowship
\#DGE-11-44155.
The first and second authors are supported in part by the Defense Advanced
Research Project Agency (DARPA) and Army Research Office (ARO) under
Contract \#W911NF-15-C-0236, and NSF grants \#CNS-1445424
and \#CCF-1423306.
Part of this work was done while the third author was a postdoctoral fellow
in the Columbia University Department of Computer Science, supported
by a junior fellowship from the Simons Society of Fellows.
Any opinions, findings and conclusions or recommendations expressed
are those of the authors and do not necessarily reflect 
the views of the the Defense Advanced Research Projects Agency, Army
Research Office, the National Science Foundation, or the
U.S.~Government. 

\fi

\ifnum\lncs=1
	\bibliographystyle{splncs03}
	\bibliography{refs}
\else
	\addcontentsline{toc}{section}{References}
	\bibliographystyle{alpha}
	\bibliography{refs}

\newcommand{\etalchar}[1]{$^{#1}$}
\begin{thebibliography}{CTUW14}

\bibitem[BDMN05]{BlumDMN05}
Avrim Blum, Cynthia Dwork, Frank McSherry, and Kobbi Nissim.
\newblock Practical privacy: the {SuLQ} framework.
\newblock In {\em {PODS}}, 2005.

\bibitem[BFM14]{BFM14}
Christina Brzuska, Pooya Farshim, and Arno Mittelbach.
\newblock Indistinguishability obfuscation and uces: The case of
  computationally unpredictable sources.
\newblock In {\em CRYPTO}, 2014.

\bibitem[BLR13]{BlumLR08}
Avrim Blum, Katrina Ligett, and Aaron Roth.
\newblock A learning theory approach to noninteractive database privacy.
\newblock {\em J. {ACM}}, 60(2):12, 2013.

\bibitem[BMSZ16]{BMSZ16}
Saikrishna Badrinarayanan, Eric Miles, Amit Sahai, and Mark Zhandry.
\newblock Post-zeroizing obfuscation: New mathematical tools, and the case of
  evasive circuits.
\newblock In {\em EUROCRYPT}, 2016.

\bibitem[BNS13]{BeimelNS13}
Amos Beimel, Kobbi Nissim, and Uri Stemmer.
\newblock Private learning and sanitization: Pure vs. approximate differential
  privacy.
\newblock In {\em {RANDOM}}, 2013.

\bibitem[BNSV15]{BunNSV15}
Mark Bun, Kobbi Nissim, Uri Stemmer, and Salil~P. Vadhan.
\newblock Differentially private release and learning of threshold functions.
\newblock In {\em {FOCS}}, 2015.

\bibitem[BPW16]{BPW16}
Nir Bitansky, Omer Paneth, and Daniel Wichs.
\newblock Perfect structure on the edge of chaos.
\newblock In {\em TCC}, 2016.

\bibitem[BST16]{BST16}
Mihir Bellare, Igors Stepanovs, and Stefano Tessaro.
\newblock Contention in cryptoland: Obfuscation, leakage and uce.
\newblock In {\em TCC}, 2016.

\bibitem[BUV14]{BunUV14}
Mark Bun, Jonathan Ullman, and Salil~P. Vadhan.
\newblock Fingerprinting codes and the price of approximate differential
  privacy.
\newblock In {\em {STOC}}, 2014.

\bibitem[BZ14]{BonehZ14}
Dan Boneh and Mark Zhandry.
\newblock Multiparty key exchange, efficient traitor tracing, and more from
  indistinguishability obfuscation.
\newblock In {\em {CRYPTO}}, 2014.

\bibitem[BZ16]{BunZ16}
Mark Bun and Mark Zhandry.
\newblock Order-revealing encryption and the hardness of private learning.
\newblock In {\em {TCC}}, 2016.

\bibitem[CFN94]{ChorFN94}
Benny Chor, Amos Fiat, and Moni Naor.
\newblock Tracing traitors.
\newblock In {\em {CRYPTO}}, pages 257--270, 1994.

\bibitem[CGH{\etalchar{+}}15]{CGHLMMRST15}
Jean-S{\'e}bastien Coron, Craig Gentry, Shai Halevi, Tancr{\`e}de Lepoint,
  Hemanta~K. Maji, Eric Miles, Mariana Raykova, Amit Sahai, and Mehdi Tibouchi.
\newblock Zeroizing without low-level zeroes: New mmap attacks and their
  limitations.
\newblock In {\em CRYPTO}, 2015.

\bibitem[CTUW14]{ChandrasekaranTUW14}
Karthekeyan Chandrasekaran, Justin Thaler, Jonathan Ullman, and Andrew Wan.
\newblock Faster private release of marginals on small databases.
\newblock In {\em Innovations in Theoretical Computer Science, ITCS'14,
  Princeton, NJ, USA, January 12-14, 2014}, pages 387--402, 2014.

\bibitem[DMNS06]{DworkMNS06}
Cynthia Dwork, Frank McSherry, Kobbi Nissim, and Adam Smith.
\newblock Calibrating noise to sensitivity in private data analysis.
\newblock In {\em {TCC}}, 2006.

\bibitem[DN03]{DinurN03}
Irit Dinur and Kobbi Nissim.
\newblock Revealing information while preserving privacy.
\newblock In {\em {PODS}}, 2003.

\bibitem[DN04]{DworkN04}
Cynthia Dwork and Kobbi Nissim.
\newblock Privacy-preserving datamining on vertically partitioned databases.
\newblock In {\em {CRYPTO}}, 2004.

\bibitem[DNR{\etalchar{+}}09]{DworkNRRV09}
Cynthia Dwork, Moni Naor, Omer Reingold, Guy~N. Rothblum, and Salil~P. Vadhan.
\newblock On the complexity of differentially private data release: efficient
  algorithms and hardness results.
\newblock In {\em {STOC}}, 2009.

\bibitem[DNT14]{DworkNT14}
Cynthia Dwork, Aleksandar Nikolov, and Kunal Talwar.
\newblock Using convex relaxations for efficiently and privately releasing
  marginals.
\newblock In {\em {SOCG}}, 2014.

\bibitem[DRV10]{DworkRV10}
Cynthia Dwork, Guy~N. Rothblum, and Salil~P. Vadhan.
\newblock Boosting and differential privacy.
\newblock In {\em {FOCS}}. {IEEE}, 2010.

\bibitem[DSS{\etalchar{+}}15]{DworkSSUV15}
Cynthia Dwork, Adam~D. Smith, Thomas Steinke, Jonathan Ullman, and Salil~P.
  Vadhan.
\newblock Robust traceability from trace amounts.
\newblock In {\em {FOCS}}, 2015.

\bibitem[GGH{\etalchar{+}}13]{GGHRSW13}
S.~Garg, C.~Gentry, S.~Halevi, M.~Raykova, A.~Sahai, and B.~Waters.
\newblock Candidate indistinguishability obfuscation and functional encryption
  for all circuits.
\newblock In {\em {FOCS}}, pages 40--49, 2013.

\bibitem[GHRU13]{GuptaHRU11}
Anupam Gupta, Moritz Hardt, Aaron Roth, and Jonathan Ullman.
\newblock Privately releasing conjunctions and the statistical query barrier.
\newblock {\em {SIAM} J. Comput.}, 42(4):1494--1520, 2013.

\bibitem[GLSW15]{GLSW15}
Craig Gentry, Allison~Bishop Lewko, Amit Sahai, and Brent Waters.
\newblock Indistinguishability obfuscation from the multilinear subgroup
  elimination assumption.
\newblock In {\em FOCS}, 2015.

\bibitem[GMS16]{GMS16}
Sanjam Garg, Pratyay Mukherjee, and Akshayaram Srinivasan.
\newblock Obfuscation without the vulnerabilities of multilinear maps.
\newblock Cryptology ePrint Archive, Report 2016/390, 2016.
\newblock \url{http://eprint.iacr.org/}.

\bibitem[GRU12]{GuptaRU12}
Anupam Gupta, Aaron Roth, and Jonathan Ullman.
\newblock Iterative constructions and private data release.
\newblock In {\em {TCC}}, 2012.

\bibitem[HR10]{HardtR10}
Moritz Hardt and Guy~N. Rothblum.
\newblock A multiplicative weights mechanism for privacy-preserving data
  analysis.
\newblock In {\em {FOCS}}, 2010.

\bibitem[HRS12]{HardtRS12}
Moritz Hardt, Guy~N. Rothblum, and Rocco~A. Servedio.
\newblock Private data release via learning thresholds.
\newblock In {\em {SODA}}, 2012.

\bibitem[HSW14]{HSW14}
Susan Hohenberger, Amit Sahai, and Brent Waters.
\newblock Replacing a random oracle: Full domain hash from indistinguishability
  obfuscation.
\newblock In {\em EUROCRYPT}, 2014.

\bibitem[HU14]{HardtU14}
Moritz Hardt and Jonathan Ullman.
\newblock Preventing false discovery in interactive data analysis is hard.
\newblock In {\em 55th {IEEE} Annual Symposium on Foundations of Computer
  Science, {FOCS} 2014, Philadelphia, PA, USA, October 18-21, 2014}, pages
  454--463, 2014.

\bibitem[Kea98]{Kearns93}
Michael~J. Kearns.
\newblock Efficient noise-tolerant learning from statistical queries.
\newblock {\em J. {ACM}}, 45(6), 1998.

\bibitem[KLN{\etalchar{+}}11]{KasiviswanathanLNRS09}
Shiva~Prasad Kasiviswanathan, Homin~K Lee, Kobbi Nissim, Sofya Raskhodnikova,
  and Adam Smith.
\newblock What can we learn privately?
\newblock {\em SIAM Journal on Computing}, 40(3):793--826, 2011.

\bibitem[MSZ16]{MSZ16}
Eric Miles, Amit Sahai, and Mark Zhandry.
\newblock Annihilation attacks for multilinear maps: Cryptanalysis of
  indistinguishability obfuscation over ggh13.
\newblock In {\em {CRYPTO}}, 2016.

\bibitem[NTZ13]{NikolovTZ13}
Aleksandar Nikolov, Kunal Talwar, and Li~Zhang.
\newblock The geometry of differential privacy: the sparse and approximate
  cases.
\newblock In {\em {STOC}}, 2013.

\bibitem[RR10]{RothR10}
Aaron Roth and Tim Roughgarden.
\newblock Interactive privacy via the median mechanism.
\newblock In {\em STOC}, pages 765--774. {ACM}, June 5--8 2010.

\bibitem[SU15a]{SteinkeU15b}
Thomas Steinke and Jonathan Ullman.
\newblock Between pure and approximate differential privacy.
\newblock {\em CoRR}, abs/1501.06095, 2015.

\bibitem[SU15b]{SteinkeU15a}
Thomas Steinke and Jonathan Ullman.
\newblock Interactive fingerprinting codes and the hardness of preventing false
  discovery.
\newblock In {\em Proceedings of The 28th Conference on Learning Theory, {COLT}
  2015, Paris, France, July 3-6, 2015}, pages 1588--1628, 2015.

\bibitem[SW14]{SW14}
Amit Sahai and Brent Waters.
\newblock How to use indistinguishability obfuscation: Deniable encryption, and
  more.
\newblock In {\em {STOC}}, 2014.

\bibitem[TUV12]{ThalerUV12}
Justin Thaler, Jonathan Ullman, and Salil~P. Vadhan.
\newblock Faster algorithms for privately releasing marginals.
\newblock In {\em {ICALP}}, 2012.

\bibitem[Ull13]{Ullman13}
Jonathan Ullman.
\newblock Answering $n^{2+o(1)}$ counting queries with differential privacy is
  hard.
\newblock In {\em {STOC}}, 2013.

\bibitem[Ull15]{Ullman15}
Jonathan Ullman.
\newblock Private multiplicative weights beyond linear queries.
\newblock In {\em {PODS}}, 2015.

\bibitem[UV11]{UllmanV11}
Jonathan Ullman and Salil~P. Vadhan.
\newblock {PCP}s and the hardness of generating private synthetic data.
\newblock In {\em {TCC}}, 2011.

\end{thebibliography}
\fi

\ifnum\lncsshort=1
\end{document}
\else

\appendix
\section{Twice Puncturable Pseudorandom Functions} \label{sec:prfconstruction}

\subsection{An Input-Matching Secure PRF}
Like pseudorandom functions satisfying the existing notions of puncturing, our construction is simply the GGM PRF family.  We detail this constraint for notational purposes.  For simplicity, we assume that $m$ and $n$ are powers of $2$.
\begin{figure}[h!t]
\begin{framed}
\begin{algorithmic}
\INDSTATE[0]{$\prf:$}
\INDSTATE[1]{Parameters: a seed $s \in \zo^{\lambda}$, a pseudorandom generator $PRG \from \zo^{\lambda} \to \zo^{2\lambda}$}
\INDSTATE[6]{~~and domain and range sizes $n = \poly(\lambda), m = \tilde{O}(n^{7})$}
\INDSTATE[1]{Input: $x \in [m] = \zo^{\mu}$ where $\mu = \log_2(m)$}
\INDSTATE[1]{Let $z = PRG_{x_{\mu}}(PRG_{x_{\mu-1}}(\ldots(PRG_{x_{2}}(PRG_{x_{1}}(s)))\dots))$}
\INDSTATE[1]{Let $y$ be the first $\log_2(n)$ bits of $z$ and output $y$}
\INDSTATE[1]{}
\INDSTATE[0]{Setup($1^{\lambda}$)}:
\INDSTATE[1]{Draw $s \from \zo^{\lambda}$ at random conditioned on every $y \in [n]$ having a preimage under the}
\INDSTATE[1]{$\prf$ defined by $s$.}
\INDSTATE[1]{Note that this is efficiently computable since $m = \omega(n \lg n)$ by the reasoning in section~\ref{def:tpprf}.}
\INDSTATE[1]{}
\INDSTATE[0]{Puncture($\prf, x_0, x_1$)}:
\INDSTATE[1]{Note that the pseudorandom function $\prf$ defines a tree of seed values}
\INDSTATE[2]{(one for each node of the tree - the root is $s$, the left child is $PRG(s)_0$, etc)}
\INDSTATE[1]{Output $\prf^{\{x_0, x_1\}} =$ the set of all seed values of each node which is not an ancestor of}
\INDSTATE[2]{$x_0$ or $x_1$ but its parent is. (note there are $O(\lg m)$ such seeds)}
\INDSTATE[0]{}
\INDSTATE[0]{$\prf^{\{x_0, x_1\}}$}
\INDSTATE[1]{Input: $x \in [m] = \zo^{\mu}$ where $\mu = \log_2(m)$}
\INDSTATE[1]{Let $s_j$ be the seed of the first node which is an ancestor of $x$ and not an ancestor of}
\INDSTATE[1]{$x_0, x_1$ but it parent is. Let $j$ be its height in the binary tree.}
\INDSTATE[1]{Let $z = PRG_{x_{\mu}}(PRG_{x_{\mu-1}}(\ldots(PRG_{x_{j+1}}(PRG_{x_{j}}(s_j)))\dots))$}
\INDSTATE[1]{Let $y$ be the first $\log_2(n)$ bits of $z$ and output $y$}
\end{algorithmic}
\end{framed}
\vspace{-6mm}
\caption{The GGM pseudorandom function family $\set{\prf: [m] \to [n]}$}
\end{figure}
We now claim that the GGM construction satisfies Theorem~\ref{thm:existstwicepuncturableprf}. 
\begin{theorem}
If $m,n$ are polynomial in the security parameter and $m = \omega(n \log(n))$, then the GGM pseudorandom function is $\eps$-input-matching secure for $\eps = \tilde{O}(\sqrt{n/m})$.
\end{theorem}

We start by modifying the {\sf InputMatching} game to one that will make it easier to prove security.  Consider the following pair of games.
\begin{itemize}
	\item {\sf Game0:} The $\mathsf{InputMatching}[y_0, y_1]$ game above.
	
	\item {\sf Game1:} We modify the $\mathsf{InputMatching}[y_0, y_1]$ game in the following way.  Instead of choosing $\prf$ conditioned on $\forall y \in [n],\, \prf^{-1}(y) \neq \emptyset$, we first choose $x_{0}, x_{1} \getsr [m]$ and then choose $\prf$ conditioned on $\prf(x_0) = y_{b_0}$ and $\prf(x_1) = y_{b_1}$.  We will prove that the challenges $(x_0, x_1, \prf^{\{x_0, x_1\}})$ in {\sf Game 0} and {\sf Game1} are statistically indistinguishable with suitable parameters.
	
\end{itemize}
\jnote{There was another game here, but for the XOR definition I'm not sure we need it.}

\begin{claim} \label{clm:game0game1}
For every polynomials $m,n$, {\sf Game0} and {\sf Game1} are $\eps$-computationally indistinguishable for $\eps = \tilde{O}(\sqrt{n/m})$.
\end{claim}
The proof is an uninsightful computation, so we will defer it to Section~\ref{sec:game0game1proof}.  Now we want to prove that the $b_0 \oplus b_1 = 0$ and $b_0 \oplus b_1 = 1$ cases of {\sf Game1} are computationally indistinguishable.
\begin{claim} \label{clm:game1}
$$
\pr{\mathsf{Game1}}{b' = b_0 \oplus b_1} \leq \frac{1}{2} + O\left(\frac{1}{m}\right).
$$
\end{claim}

To show this, we need the following lemma:

\begin{lemma}\label{lem:conditionedprg} Let $\prg:\zo^{\lambda} \rightarrow \zo^{2\lambda}$ be a PRG, and let $p:\zo^{2\lambda}\rightarrow\{0,1\}$ be an efficiently computable predicate on $\zo^{2\lambda}$.  Suppose that, for a random $z\in\zo^{2\lambda}$, $\Pr[p(z)=1]$ is non-negligible.  Then the following distributions are $\eps$-computationally indistinguishable for $\eps = \negl(\lambda)$.
	\begin{itemize}
		\item $z$ for $z=\prg(s)$,where $s$ is chosen at random from $\zo^{\lambda}$ conditioned on $p(\prg(s))=1$.
		\item $z$ where $z$ is chosen at random from $\zo^{2\lambda}$ conditioned on $p(z)=1$.
	\end{itemize}
\end{lemma}
\begin{proof}[Proof sketch] Given an efficient distinguisher $D$ for the two distributions above, we construct the following efficient $\prg$ distinguisher $D'$.  $D'$, on input $z$, computes $p(z)$.  Since $p$ and $D$ are computationally efficient, so is $D'$.  If the output is 0, then $D'$ outputs a random bit.  Otherwise, $D'$ runs $D$ on $z$ and outputs the result.  In the case where $p(z)=0$, $D'$ has no advantage, and in the case where $p(z)=1$, $D'$ has non-negligible advantage (namely the advantage of $D$).  Since $p(z)=1$ with non-negligible probability, $D'$ has overall non-negligible advantage.
\end{proof}

	\jnote{So the predicate here is $ p ={ }$[the value $s$ is consistent with the values of the $\prf$ on $x_0, x_1$]?}
	\mnote{Yes, that is correct}

\begin{proof}[Proof sketch of Claim~\ref{clm:game1}]
Consider the GGM tree, and the punctured function $\prf^{\{x_0,x_1\}}$ consisting of the values at all nodes in the tree that are not an ancestor of $x_0$ or $x_1$, but whose parent is an ancestor.  \jnote{Need to define the puncturing algorithm, but once we do I think this level of detail is OK.} We now consider the following procedure
\begin{enumerate}
	\item Pick a node whose value $s$ is random (perhaps conditioned on some predicate $p$ on $\prg(s)$), and not derived from another node's value.  For example, at the beginning this is the root node, which is random, conditioned on the leaves at $x_0$ and $x_1$ having values $y_{b_{0}},y_{b_{1}}$.
	\item If that node is part of the punctured key $\prf^{\{x_0,x_1\}}$ or is one of the leaves at $x_0,x_1$, don't do anything to this node.
	\item Otherwise, delete that node, and replace the values of the children with random, conditioned on the predicate $p$ being 1.  This predicate is applied to the concatenation of the children's values.  Notice that, given the form of our initial predicate each of the children's new values $s_0,s_1$ will be independently random, perhaps conditioned on some predicates $p_0,p_1$ on $\prg(s_0),\prg(s_1)$ respectively.  
\end{enumerate}
We iterate the procedure until we no longer make any changes to the tree.  By applying Lemma~\ref{lem:conditionedprg} to the change made in step 3, we can see that the distributions on the punctured PRF before and after we apply the procedure are $\eps$-computationally indistinguishable for $\eps = \negl(\lambda)$.  In the end, we will have changed all of the punctured key values to uniformly random.  Note that, since these nodes are not ancestors of $x_0,x_1$, the predicate on them is trivially satisfied, so they are uniformly random even conditioned on the values of the function at $x_0,x_1$.  Thus, these values are also independent of the output of the function at the points $x_0, x_1$.  Thus, as long as $x_0 \neq x_1$ we can swap these values to any combination of $y_{b_{0}}$ and $y_{b_{1}}$.  Since the probability that $x_0 = x_1$ is at most $1/m$, the probability that the adversary can guess $b_{0} \oplus b_{1}$ is at most $1/2 + 1/m$.  Since this distribution is $\eps$-computationally indistinguishable from the real distribution on the punctured PRF, a computationally efficient adversary can guess $b_0 \oplus b_1$ with probability at most $1/2 + 1/m + \negl(\lambda) = 1/2 + O(1/m)$.
\end{proof}

\subsubsection{Proof of Claim~\ref{clm:game0game1}} \label{sec:game0game1proof}
Consider two ways of sampling a tuple $(y_0, y_1, b_0, b_1, x_0, x_1, \prf)$.  Here $y_0, y_1 \in [n]$ are fixed, $b_0, b_1 \in \zo$ are uniformly random and independent.  In {\sf Game0}, $\prf$ is sampled conditioned on every $y \in [n]$ having non-empty preimage, and $x_0$ and $x_1$ are random preimages of $y_{b_0}$ and $y_{b_1}$, respectively.  In {\sf Game1}, $x_0, x_1 \in [m]$ are uniformly random and independent and $\prf$ is sampled conditioned on $\prf(x_{0}) = y_{b_{0}}$ and $\prf(x_{1}) = y_{b_{1}}$.  Observe that these tuples contain strictly more information than the challenges given to the adversary in {\sf Game0} and {\sf Game1}, respectively.  That is, the challenges can be generated by applying a function to these tuples, which cannot increase the statistical distance between the two distributions.  Thus, to prove Claim~\ref{clm:game0game1} it suffices to prove that these two distributions are statistically close.

First, we switch to an intermediate game {\sf Game0A} in which the function $\prf$ is \emph{not} required to have a preimage for every $y \in [n]$.  To make the sampling procedure well defined, if $y_{b_{0}}$ or $y_{b_{1}}$ does not have a preimage under $\prf$, we simply choose $x_0$ and $x_1$ to be $\bot$.
\begin{lemma} \label{lem:game0game0A}
If $\prf$ is pseudorandom and $m, n$ are polynomials, then {\sf Game0} and {\sf Game0A} are $\eps$-statistically indistinguishable for $\eps = n\cdot \exp(-\Omega(m/n)) + \negl(\lambda)$.
\end{lemma}
\begin{proof}
Suppose that $f \from [m] \to [n]$ is a uniformly random function.  Then a simple calculation shows that the probability that there exists $y \in [n]$ such that $\prf^{-1}(y) = \emptyset$ is at most
$$
n \cdot (1-1/n)^{m} = n \cdot \exp(-\Omega(m/n)).
$$
Since $m,n$ are polynomial in the security parameter $\lambda$, there is a polynomial time algorithm that checks whether a function $\prf \from [m] \to [n]$ has at least one preimage for every $y \in [n]$.  Thus, if $\prf$ is sampled from a pseudorandom function family, it must also be true that the probability that there exists $y \in [n]$ such that $\prf^{-1}(y) = \emptyset$ is at most $n \cdot \exp(-\Omega(m/n)) + \negl(\lambda)$, or else there would be an efficient algorithm that distinguishes a random function $\prf$ from a truly random function $f$.

Since conditioning on an event that occurs with probability at least $1-p$ can only affect the distribution by at most $p$ in statistical distance, we conclude that the two distributions are statistically indistinguishable to within $n \cdot \exp(-\Omega(m/n)) + \negl(\lambda)$.
\end{proof}

Now, we introduce a second intermediate game {\sf Game0B} in which we first choose a random value $x_0 \in [m]$, then sample $\prf$ such that $\prf(x_0) = y_{b_{0}}$, and finally we choose $x_1$ to be a random preimage of $y_{b_{1}}$.  If $y_{b_{1}}$ has no preimage under $\prf$, we set $x_1 = \bot$.
\begin{lemma} \label{lem:game0Agame0B}
If $\prf$ is pseudorandom and $m, n$ are polynomials, then {\sf Game0A} and {\sf Game0B} are $\eps$-statistically indistinguishable for $\eps = \tilde{O}(\sqrt{n/m})$.
\end{lemma}

Before proving the lemma, we will state and prove a useful combinatorial lemma about conditioning a pseudorandom function on a single input-output pair.  Consider the following two ways of sampling a pseudorandom function.
\begin{figure}[h!t]
\begin{framed}
\begin{algorithmic}
\INDSTATE[0]{Choose a pseudorandom function $\prf: [m] \to [n]$}
\INDSTATE[0]{Let $s = |\prf^{-1}(y)|$.}
\INDSTATE[0]{If $s = 0$, let $x = \bot$, else choose a random $x \in [m]$ from $\prf^{-1}(y)$.}
\INDSTATE[0]{Output $(x, \prf)$.}
\end{algorithmic}
\end{framed}
\vspace{-6mm}
\caption{$\mathsf{ExpA}[y]$}
\end{figure}
\begin{figure}[h!t]
\begin{framed}
\begin{algorithmic}
\INDSTATE[0]{Choose a random $x \in [m]$}
\INDSTATE[0]{Choose a pseudorandom function $\prf: [m] \to [n]$ so that $\prf(x) = y$.}
\INDSTATE[0]{Let $s = |\prf^{-1}(y)|$.}
\INDSTATE[0]{Output $(x, \prf)$}
\end{algorithmic}
\end{framed}
\vspace{-6mm}
\caption{$\mathsf{ExpB}[y]$}
\end{figure}
\begin{lemma} \label{lem:stupidcombinatorics}
If $\prf$ is pseudorandom, and $m,n$ are polynomials, then for every $y \in [n]$, $\mathsf{ExpA}[y]$ and $\mathsf{ExpB}[y]$ are $\tilde{O}(\sqrt{n/m}) + \negl(\lambda)$ computationally indistinguishable.
\end{lemma}
\begin{proof}[Proof of Lemma~\ref{lem:stupidcombinatorics}]
First, we will replace $\prf \from [m] \to [n]$ with a truly random function $f \from [m] \to [n]$.  We will argue later why using a pseudorandom function cannot increase the statistical distance between the two distributions by more than $\negl(\lambda)$.

Now, observe that in both experiments, the marginal distribution on $x$ is uniform on $[m]$.  Also, observe that for every fixed choice of $x$ and $s$, the conditional distribution $f | x,s$ in each experiment is the same.  Finally, note that the distribution on $s$ is independent of $x$ in each experiment.  Thus, in order to bound the statistical distance between the two experiments, it suffices to bound the statistical distance between the marginal distributions of $s$ in the two experiments.

\biglnote{Isn't the marginal distribution on $x$ in ExpA not uniform? (It can be bottom with probability $\frac{1}{m^{n}}$ in this experiment).}

In $\mathsf{ExpA}$, the probability that $|f^{-1}(y)| = s$ is precisely the probability that $y$ has exactly $s$ preimages in a random function $f : [m] \to [n]$
\begin{align*}
\pr{f: [m] \to [n]}{|f^{-1}(y)| = s}
={} \binom{m}{s} \left( \frac{1}{n} \right)^{s} \left( 1 - \frac{1}{n} \right)^{m-s}
\end{align*}

In $\mathsf{ExpB}$, the probability that $|f^{-1}(y)| = s$ is precisely the probability that $y$ has exactly $s-1$ preimages in a random function $f : [m-1] \to [n]$, since we fix the fact that $f(x) = y$ and the remainder of the function is chosen randomly.
\begin{align*}
&\pr{f: [m-1] \to [n]}{|f^{-1}(y)| = s-1}
={} \binom{m-1}{s-1} \left( \frac{1}{n} \right)^{s-1} \left( 1 - \frac{1}{n} \right)^{m-s}
\end{align*}

Thus, the statistical distance between the two distributions is
\begin{align*}
&\sum_{s = 1}^{m} \left| \binom{m}{s} \left( \frac{1}{n} \right)^{s} \left( 1 - \frac{1}{n} \right)^{m-s} - \binom{m-1}{s-1} \left( \frac{1}{n} \right)^{s-1} \left( 1 - \frac{1}{n} \right)^{m-s} \right| \\
={} &\sum_{s = 1}^{m} \left(\frac{1}{n}\right)^{s-1} \left( 1 - \frac{1}{n} \right)^{m-s} \left| \binom{m}{s} \left( \frac{1}{n} \right) - \binom{m-1}{s-1}  \right| \\
={} &\sum_{s = 1}^{m} \binom{m-1}{s-1} \left(\frac{1}{n}\right)^{s-1} \left( 1 - \frac{1}{n} \right)^{m-s} \left| \frac{m}{sn} - 1  \right|
\end{align*}
To tackle the final sum, we consider two cases roughly corresponding to whether $m/sn - 1$ is close to $0$ or far from $0$.  Typically, when we choose a random function from either distribution we will have a preimage size of $s \approx m/n$, in which case $|m/sn - 1| \approx 0$.  Since, by the binomial theorem, $$\sum_{s=1}^{m} \binom{m-1}{s-1} (1/n)^{s-1} (1 - 1/n)^{m-s} = \sum_{s = 0}^{m-1} \binom{m-1}{s} (1/n)^s (1 - 1/n)^{m-1-s} = 1,$$ we will have that this portion of the sum is close to $0$.  In the atypical case, we will have that $s$ is far from $m/n$, in this case we will use the fact that the probability of choosing a random function $f$ with preimage size far from $m/n$ is much smaller than $n/m$, and $|m/sn - 1| \leq m/n$, to conclude that this portion of the sum is also close to $0$.

Specifically, fix some threshold $\tau$ and break the sum into two regions based on whether or not $s \in (1 \pm \tau) m/n$.
\begin{align*}
&\sum_{s = 1}^{m} \binom{m-1}{s-1} \left(\frac{1}{n}\right)^{s-1} \left( 1 - \frac{1}{n} \right)^{m-s} \left| \frac{m}{sn} - 1  \right| \\
={} &\sum_{s \in (1\pm \tau) m/n } \binom{m-1}{s-1} \left(\frac{1}{n}\right)^{s-1} \left( 1 - \frac{1}{n} \right)^{m-s} \left| \frac{m}{sn} - 1  \right| + \sum_{s \not\in (1 \pm \tau)m/n} \binom{m-1}{s-1} \left(\frac{1}{n}\right)^{s-1} \left( 1 - \frac{1}{n} \right)^{m-s} \left| \frac{m}{sn} - 1  \right| \\
\leq{} &2\tau \cdot \sum_{s \in (1 \pm \tau) m/n } \binom{m-1}{s-1} \left(\frac{1}{n}\right)^{s-1} \left( 1 - \frac{1}{n} \right)^{m-s} + \frac{m}{n} \cdot \sum_{s \not\in (1 \pm \tau) m/n} \binom{m-1}{s-1} \left(\frac{1}{n}\right)^{s-1} \left( 1 - \frac{1}{n} \right)^{m-s} \\
\leq{} &2\tau + \frac{m}{n} \cdot \pr{f: [m-1] \to [n]}{|f^{-1}(y)| \not\in (1 \pm \tau) m/n} \tag{Definition of Binomial Distribution} \\
\leq{} &2\tau + \frac{m}{n} \cdot e^{-\Omega(\tau^2 m / n)} \tag{Chernoff bound}.
\end{align*}
In this calculation, we make use of a form of the Chernoff bound that states if $X_1,\dots,X_T$ are independent random variables taking values in $\zo$, and $X = \sum_{t=1}^{T} X_t$, then for every $\tau <= 1$, $\pr{}{X \not\in (1 \pm \tau)\ex{}{X}} \leq e^{-\Omega(\tau^2 \ex{}{X})}$.  

From this calculation, it is clear that there is a setting of $\tau = \tilde{O}(\sqrt{n/m})$ such that the final expression is bounded by $O(\tau) = \tilde{O}(\sqrt{n/m})$.  Putting it together, the statistical distance between the two distributions in question is $\tilde{O}(\sqrt{n/m}) + e^{-\Omega(m/n)} = \tilde{O}(\sqrt{n/m})$.  This completes the proof.

\jnote{This last paragraph is super sketchy.}
Finally, we have to argue that the two distributions remain close if we use a pseudorandom function in place of a truly random function.  Since $m,n$ are polynomial, an efficient adversary can enumerate all of the input-output pairs of the function $\prf$.  Thus, in order for $\prf$ to be pseudorandom, the truth table of a random $\prf$ must be computationally indistinguishable from the truth table of a random function.  By the above analysis, the distribution of the truth table of $f$ depends only on $x$ and $s$.  Thus, for every fixed value of $x,s$, it must be that the distribution of $\prf \mid x,s$ and the distribution of $f \mid x,s$ are $\eps_{x,s}$ computationally indistinguishable for $\eps_{x,s} = \negl(\lambda) / \pr{}{x,s}$.  Moreover, given $(x, \prf)$, $s$ is efficiently computable because $m,n$ are polynomial.  Thus, the two distributions are $\eps$-computationally indistinguishable for some $\eps \leq \sum_{x,s} \pr{}{x,s} (\negl(\lambda)/\pr{}{x,s}) = mn \cdot \negl(\lambda) = \negl(\lambda)$.

Putting it together, we have that {\sf ExpA} and {\sf ExpB} are $\eps$-computationally indistinguishable for $\eps = \tilde{O}(\sqrt{n/m}) + \negl(\lambda)$, as desired.
\end{proof}

Now we return to proving Lemma~\ref{lem:game0Agame0B}
\begin{proof}[Proof of Lemma~\ref{lem:game0Agame0B}]
First, fix any choice of $(y_0, y_1, b_0, b_1)$.  We want to show that the distributions on $(x_0, x_1, \prf)$ in {\sf Game0A} and {\sf Game0B} are close.  First consider just the distribution on $(x_0, \prf) \mid (y_0, y_1, b_0, b_1)$.  In {\sf Game0A}, this distribution is exactly $\mathsf{ExpA}[y_{b_{0}}]$.  In ${\sf Game0B}$, this distribution is exactly $\mathsf{ExpB}[y_{b_{0}}]$.  Thus, by Lemma~\ref{lem:stupidcombinatorics}, the two distributions are $\eps$-computationally indistinguishable for $\eps = \tilde{O}(\sqrt{n/m}) + \negl(\lambda)$.

Now, we consider the case where $b_1 = b_0$.  In this case, in both {\sf Game0A} and {\sf Game0B}, $x_{1}$ is a random preimage of $y_{b_{1}} = y_{b_{0}}$.  Unless $x_1 = x_0$, $x_1$ is a uniformly random value in $[m]$.  Since the collision probability is determined only by the number of preimages of $y_{b_{0}}$, Lemma~\ref{lem:stupidcombinatorics} also shows that the distribution on $x_{1} \mid (y_0, y_1, b_0, b_1, x_0, \prf)$ is $\eps$-computationally indistinguishable for $\eps = \tilde{O}(\sqrt{n/m} + \negl(\lambda))$.

Now, we consider the case where $b_1 \neq b_0$.  In this case the two values never collide, and once again the distribution of $x_{1} \mid (y_0, y_1, b_0, b_1, x_0, \prf)$ is determined only by the number of preimages of $y_{b_{1}}$.  Thus, we can again apply Lemma~\ref{lem:stupidcombinatorics} to argue that these two distributions are $\eps$-computationally indistinguishable.  This completes the proof of the Lemma.
\end{proof}

We can now state and prove the final step of the hybrid argument.
\begin{lemma} \label{lem:game0Bgame1}
If $\prf$ is pseudorandom and $m, n$ are polynomials, then {\sf Game0B} and {\sf Game1} are $\eps$-computationally indistinguishable for $\eps = \tilde{O}(\sqrt{n/m})$.
\end{lemma}
\begin{proof}
In either game, $x_1$ is uniformly random in $[m]$ if we condition on the event that $x_0 \neq x_1$.  First consider the case where $b_0 = b_1$, then in {\sf Game0B} then the probability of a collision is determined by the number of preimages of $y_{b_{1}} = y_{b_{0}}$, and by Lemma~\ref{lem:stupidcombinatorics} the collision probability is at most $\eps = \tilde{O}(\sqrt{n/m}) + \negl(\lambda)$.  However, in {\sf Game1} the probability of collision is exactly $1/m$.  Now, in the case where $b_0 \neq b_1$, in {\sf Game0B} the probability of collision is $0$, whereas in {\sf Game1} the probability of collision is exactly $1/m$.  Putting it together completes the proof of the Lemma.
\end{proof}

Combining the hybrids in Lemmata~\ref{lem:game0game0A},~\ref{lem:game0Agame0B},and~\ref{lem:game0Bgame1} completes the proof of Claim~\ref{clm:game0game1}.  We remark that since $m$ and $n$ are polynomials in $\lambda$, the $\negl(\lambda)$ term is of a lower order than $\tilde{O}(\sqrt{n/m})$ so we are justified in dropping it from the asymptotic expression for the distinguishing probability.

\end{document}
\fi

\subsection{Proof of Lemma~\ref{lem:conditionalhashlemma}} \label{sec:proofofconditionalhashlemma}

We will fix $b = 1$ for simplicity.  The case of $b = 0$ follows symmetrically.

We will first bound the R\'{e}nyi divergence between $D_1$ and $D_2$, which is defined as
Define the R\'{e}nyi divergence between $D_2$ and $D_1$ to be
$$
RD(D_1,D_2)=\sum_\hash \frac{\pr{}{\hash:\hash \getsr D_2}^2}{\pr{}{\hash:\hash \getsr D_1}}
$$
Here, $\hash$ ranges over the support of $D_1$.  Once we do this, we will obtain an upper bound on the statistical distance between $D_1$ and $D_2$ using the inequality
$$
SD(D_1,D_2)\leq\frac{\sqrt{RD(D_1,D_2)-1}}{2}.
$$

To bound the R\'{e}nyi divergence, we can start by writing
\begin{align*}
\pr{}{\hash:\hash \gets D_2}^2&=\left(\frac{1}{m}\sum_{x \in M}\pr{}{\hash:\hash(x)=1}\right)^2\\
&= \frac{1}{m^2}\sum_{x,x' \in M}\pr{}{\hash:\hash(x)=1}\pr{}{\hash:\hash(x')=1}
\end{align*}
Where in all the (conditional) probabilities on the right, $\hash$ is drawn from $D_1$, conditioned on some event.  
This allows us to write
$$
RD(D_1,D_2)=\frac{1}{m^2}\sum_{x,x' \in M}\sum_\hash \frac{\Pr[H=\hash:H(x)=1]\Pr[H=\hash:H(x')=1]}{\Pr[H=\hash]}
$$

We now divide the sum into two cases.
\begin{itemize}
	\item $x=x'$.  In this case, the summand becomes $\Pr[H=\hash:H(x)=1]^2/\Pr[H=\hash]$.  Notice that \[\Pr[H=\hash:H(x)=1]=\begin{cases}0&\text{if }f(x)=0\\\frac{\Pr[H=\hash]}{\Pr[H(x)=1]}\end{cases}\]
		
	Therefore, the summand is \[\frac{\Pr[H=\hash:H(x)=1]^2}{\Pr[H=\hash]}=\begin{cases}0&\text{if }f(x)=0\\\frac{\Pr[H=\hash]}{\Pr[H(x)=1]^2}&\text{if }f(x)=1\end{cases}\]
	
	Now, notice that $\sum_{f:f(x)=1}\Pr[H=\hash]=\Pr[H(x)=1]$.  Thus, if we carry out the sum over $f$, the summand becomes $1/\Pr[H(x)=1]$.  
	
	\item $x\neq x'$.  Then 
	\begin{align*}
	\Pr[H=\hash:H(x)=1]=&\Pr[H=\hash:H(x)=1,\prf(x')=0]\Pr[\prf(x')=0:H(x)=1]+\\&\Pr[H=\hash:H(x)=1,H(x')=1]\Pr[H(x')=1:H(x)=1]\\
	\Pr[H=\hash:H(x')=1]=&\Pr[H=\hash:H(x')=1,\prf(x)=0]\Pr[\prf(x)=0:H(x')=1]+\\&\Pr[H=\hash:H(x')=1,H(x)=1]\Pr[H(x)=1:H(x')=1]
	\end{align*}
	
	When we take the product of the two expressions and expand, we obtain four products of the form 
	\begin{align*}
	&\Pr[H=\hash:\prf(x)=a,\prf(x')=b]\Pr[H=\hash:\prf(x)=c,\prf(x')=d]\times\\
	&\;\;\;\;\Pr[\prf(x')=b:\prf(x)=a]\Pr[\prf(x)=c:\prf(x')=d]
	\end{align*}
	as $(a,b,c,d)$ ranges over $(1,1,1,1),(1,1,0,1),(1,0,1,1),(1,0,0,1)$.  Notice, however, that for any $f$, the product is non-zero only if $a=f(x)=c$ and $b=f(x')=d$.  Only the product $a=b=c=d=1$ satisfies this.  
	
	Therefore, the summand is 
	\begin{multline*}
	\frac{\Pr[H=\hash:H(x)=1]\Pr[H=\hash:H(x')=1]}{\Pr[H=\hash]}\\
	=\frac{\Pr[H(x')=1:H(x)=1]\Pr[H(x)=1:H(x')=1]}{\Pr[H=\hash]}\Pr[H=\hash:H(x)=1,H(x')=1]^2\\
	=\frac{\Pr[H(x')=1:H(x)=1]\Pr[H(x)=1:H(x')=1]}{\Pr[H=\hash]}\begin{cases}
	0&\text{if }f(x)=0\text{ or }f(x')=0\\
	\left(\frac{\Pr[H=\hash]}{\Pr[H(x)=1\wedgeH(x')=1]}\right)^2&\text{if }f(x)=f(x')=1
	\end{cases}\\
	=\frac{1}{\Pr[H(x)=1]\Pr[H(x')=1]}\begin{cases}
	0&\text{if }f(x)=0\text{ or }f(x')=0\\
	\Pr[H=\hash]&\text{if }f(x)=f(x')=1
	\end{cases}
	\end{multline*}
	
	When we sum over all $f$, we get \[\frac{\Pr[H=\hash:H(x)=1]\Pr[H=\hash:H(x')=1]}{\Pr[H=\hash]}=\frac{\Pr[H(x)=1\wedgeH(x')=1]}{\Pr[H(x)=1]\Pr[H(x')=1]}\]
\end{itemize}

Therefore, the R\'{e}nyi divergence is
\[RD(D_1,D_2)=\frac{1}{m^2}\left(\left(\sum_x \frac{1}{\Pr[H(x)=1]}\right)+\left(\sum_{x\neq x'}\frac{\Pr[H(x)=1\wedgeH(x')=1]}{\Pr[H(x)=1]\Pr[H(x')=1]}\right)\right)\]

We now invoke the security of PRF to see that:
\begin{itemize}
	\item For each $x\in[m]$, $\left|\Pr[H(x)=1]-\frac{1}{2}\right|\leq \negl$.  This follows by considering the simple PRF adversary that just queries on $x$ and outputs the result.  
	\item For each $x\neq x'$, $\left|\Pr[H(x)=1\wedgeH(x')=1]-\frac{1}{4}\right|\leq \negl$.  This follows by considering the PRF adversary that queries on $x,x'$, and outputs the AND of the results.
\end{itemize}

Therefore, $1/\Pr[H(x)=1]\leq 2+\negl$ and $\frac{\Pr[H(x)=1\wedgeH(x')=1]}{\Pr[H(x)=1]\Pr[H(x')=1]}\leq 1+\negl$.

Therefore, we have that \[RD(D_1,D_2)\leq \frac{1}{m^2}\left(m(2+\negl)+(m^2-m)(1+\negl)\right)=1+\frac{1}{m}+\negl\]

Next, we use the fact that the statistical distance $\Delta(D_1,D_2)$ between $D_1$ and $D_2$ satisfies \[\Delta(D_1,D_2)\leq\frac{1}{2}\sqrt{RD(D_1,D_2)-1}\leq \frac{1}{2\sqrt{m}}+\negl\]